\documentclass{aa}
\usepackage{amsmath}
\usepackage{graphicx}
\usepackage{txfonts}
\usepackage[authoryear]{natbib}
\usepackage{booktabs}
\bibpunct{(}{)}{;}{a}{}{,}

\begin{document}

\title{On the origin of LS~5039 and PSR~J1825$-$1446}


\author{J. Mold\'on\inst{\ref{inst1}}\and
M. Rib\'o\inst{\ref{inst1}}\and
J.M. Paredes\inst{\ref{inst1}}\and
W. Brisken\inst{\ref{inst2}}\and
V. Dhawan\inst{\ref{inst2}}\and
M. Kramer\inst{\ref{inst3},\ref{inst4}}\and
A.G. Lyne\inst{\ref{inst4}}\and
B.W. Stappers\inst{\ref{inst4}}}

\institute{Departament d'Astronomia i Meteorologia, Institut de Ci\`encies del
Cosmos (ICC), Universitat de Barcelona (IEEC-UB), Mart\'{\i} i Franqu\`es 1,
08028 Barcelona, Spain\label{inst1}\\ \email{jmoldon@am.ub.es}\and
National Radio Astronomy Observatory, P.O. Box 0, Socorro, NM 87801, USA\label{inst2}\and
Max Planck Institut f\"ur Radioastronomie, Auf dem Huegel 69, 53121 Bonn, Germany \label{inst3}\and
Jodrell Bank Centre for Astrophysics, School of Physics and Astronomy, University of Manchester, Manchester M13 9PL, UK\label{inst4}}

\authorrunning{Mold\'on et~al.}
\titlerunning{On the origin of LS~5039 and PSR~J1825$-$1446}

\date{Received / Accepted}

\abstract
{The gamma-ray binary LS~5039 and the isolated pulsar PSR~J1825$-$1446 were
proposed to have been formed in the supernova remnant (SNR)~G016.8$-$01.1.}
{We aim to obtain the Galactic trajectory of LS~5039 and PSR~J1825$-$1446 to find their
origin in the Galaxy, and in particular to check their association with
SNR~G016.8$-$01.1 to restrict their age.}
{By means of radio and optical observations we obtained the proper motion and the
space velocity of the sources.}
{The proper motion of PSR~J1825$-$1446 corresponds to a transverse space
velocity of 690~km~s$^{-1}$ at a distance of 5~kpc. Its Galactic velocity at
different distances is not compatible with the expected Galactic rotation. The
velocity and characteristic age of PSR~J1825$-$1446 make it incompatible with
SNR~G016.8$-$01.1. There are no clear OB associations or SNRs
crossing the past trajectory of PSR~J1825$-$1446. We estimate the age of the
pulsar to be 80--245~kyr, which is compatible with its characteristic age. The proper
motion of LS~5039 is $\mu_{\alpha}\cos\delta = 7.09$ and $\mu_{\delta} =
-8.82$~{mas~yr}$^{-1}$. The association of LS~5039 with SNR~G016.8$-$01.1 is
unlikely, although we cannot to discard it. The system would have had to be
formed in the association Ser~OB2 (at 2.0~kpc) if the age of the system is
1.0--1.2~Myr, or in the association Sct~OB3 (distance 1.5--2~kpc) for an age of
0.1--0.2~Myr. If the system were not formed close to Ser~OB2, the
pseudo-synchronization of the orbit would be unlikely.}
{PSR~J1825$-$1446 is a high-velocity isolated pulsar ejected from the Galaxy.
The distance to LS~5039, which needs to be constrained by future astrometric missions such
as Gaia, is a key parameter for restricting its origin and age.}


\maketitle

\section{Introduction} \label{introduction}

During a core-collapse supernova explosion of a massive star, the remaining
compact object, either isolated or in a binary system, can receive a natal kick
and obtain a high peculiar velocity \citep{hills83,heuvel00}. A binary system
can obtain a moderate velocity from a symmetric Blaauw kick \citep{blaauw61},
whereas the compact object can acquire additional momentum from an asymmetric
supernova kick \citep{stone82}. If the system remains bound after the explosion,
the parameters of the binary system before and after the supernova explosion can
be related by measuring the high peculiar velocity of the system, in particular
for X-ray binaries (see e.g. \citealt{brandt95}, and the particular cases in
\citealt{tauris99,martin09}). On the other hand, the space velocity of a binary
system can be used to determine its past Galactic trajectory \citep[see
e.g.][]{ribo02, dhawan07, miller-jones09}. Of particular interest are the
runaway systems, which have acquired very high peculiar velocities
\citep{mirabel01, mirabel02}. With enough information, a complete evolutionary
history of the binary system can be obtained with constraints on the progenitor and the
formation of the compact object, \citep[see
e.g.][]{willems05,fragos09}. Among the binary systems receiving natal kicks we
can find the special case of gamma-ray binaries, which usually form eccentric
binary systems \citep{casares12hess}.

Only a few binary systems have been associated with gamma-ray sources, and therefore
it is important to understand how they formed, and how they are distributed
in the Galaxy. However, the origin of most of the binaries with gamma-ray
emission is still unknown. \cite{mirabel04} studied the origin of the gamma-ray
binary \object{LS~I~+61~303}, and \cite{dhawan06} measured a precise proper
motion that allowed the authors to obtain the space velocity of the source and
constrain the mass lost in the supernova event that formed the compact object of
the system to $\sim$1~M$_{\odot}$. The natal kick of the supernova that produced
the pulsar in the system \object{PSR~B1259$-$63}/\object{LS~2883} was discussed
in \cite{hughes99} and \cite{wang06}, although no proper motion has been
measured for this system. The Galactic motion of \object{LS~5039} was
extensively discussed in \cite{ribo02} (see details in Sect.~\ref{ls_snr}).
\cite{mirabel03} and \cite{reid11} used high-accuracy radio astrometry to
compute the Galactic velocity of \object{Cygnus~X-1}, which allowed them to
obtain the mass of the progenitor of the black hole, and its association with
the Cyg~OB2 association. They also determined that the black hole was formed
without a supernova explosion. For \object{Cygnus~X-3},
\cite{miller-jones09a} obtained an accurate proper motion of the source that was
used to better identify the position of the X-ray binary core, although the
origin of the system is still unknown. \cite{moldon11_hess} determined for
\object{HESS~J0632+057} that its total proper motion is below
4~mas~yr$^{-1}$. Other recent gamma-ray binary candidates are
\object{1FGL~J1018.6$-$5856} \citep{corbet11,pavlov11,ackermann12} and
\object{AGL~J2241+4454} \citep{williams10, casares12hess}. However, the proper
motion of these sources is still unknown. These examples show the potential of a
good determination of the proper motion of a system, and also the need for better
astrometry in this peculiar binary population.

On the other hand, accurate astrometry of isolated pulsars opens a wide field of
scientific research including fundamental reference frame ties, physics in the
core-collapse supernovae and imparted momentum kicks, association with supernova
remnants (SNRs) and determination of the age of pulsars, model-independent estimates of
distances through parallax measurements, or determination of the distribution of
electron density in the interstellar medium (see for example
\citealt{chatterjee09,brisken02} and references therein). It is common that
pulsars acquire high velocities at birth, becoming the fastest population in the
Galaxy, with a mean transverse velocity of $\sim450$~km~s$^{-1}$, and up to
above $10^{3}$~km~s$^{-1}$ \citep{lyne94,hobbs05, chatterjee05}. The pulsar
velocity distribution provides information about the supernova symmetry and the
binary population synthesis \citep{brisken03}.

Here we present the results from two high-resolution astrometric projects to
determine the proper motion of \object{LS~5039}, and the nearby pulsar
\object{PSR~J1825$-$1446}, the only known compact objects in the field of the
\object{SNR~G016.8$-$01.1}. In Sect.~\ref{context} we describe
the gamma-ray binary \object{LS~5039}, its possible association with
\object{SNR~G016.8$-$01.1}, and the motivation to obtain the proper motion of
the pulsar \object{PSR~J1825$-$1446}. In Sect.~\ref{obs_psr} we describe the
VLBA astrometric project on \object{PSR~J1825$-$1446}, and we obtain the proper
motion of the source. In Sect.~\ref{obs_ls5039} we present three sets of radio
observations of \object{LS~5039}, as well as the available radio and optical
astrometry of the source, and we compute the proper motion and discuss the
uncertainties of the fit. In Sect.~\ref{gal_vel} we compute the Galactic
velocity of both sources. In Sect.~\ref{origin} we analyse the past trajectories
of the sources and discuss their possible birth location. Finally, we summarise
the obtained results in Sect.~\ref{conclusions}.

\section{Compact objects in the field of SNR~G016.8$-$01.1} \label{context}

\subsection{The gamma-ray binary LS~5039} \label{ls_snr}

\object{LS~5039} is a gamma-ray binary system that displays non-thermal
persistent and variable emission from radio frequencies to high-energy (HE;
$E>100$~MeV) and very-high-energy (VHE; $E>100$~GeV) gamma rays. The system
contains a bright ON6.5\,V((f)) star \citep{clark01,mcswain04} and a compact
object of unknown nature. This degenerate companion has a mass greater than
$\sim1.5$~M$_\odot$, which depends on the binary system inclination, which is an
unknown parameter \citep{casares05}. The orbit of the system has a period of
3.9~days and an eccentricity around 0.35 \citep{casares05,aragona09, sarty11}.
The distance to the source has recently been updated to $2.9\pm0.8$~kpc \citep{casares12_ls}.
The persistent synchrotron radio emission
\citep{marti98,ribo99,godambe08,bhattacharyya12} appears extended when observed
Baseline Interferometry (VLBI), at scales of 5--300~milliarcsecond (mas)
\citep{paredes00, paredes02,ribo08}. The X-ray spectrum of the system
\citep{reig03,martocchia05} is well fitted by a simple power law, and it is
clearly periodic, as shown by {\it RXTE} and \emph{Suzaku} \citep{bosch-ramon05,
takahashi09}. \cite{durant11} discovered a large-scale extended component in
X-rays up to 1$^{\prime}$ from \object{LS~5039}. The system has also been
detected by \emph{INTEGRAL} up to 200~keV \citep{goldoni07,hoffmann09}, and with
BATSE up to 1~MeV \citep{harmon04}. \object{LS~5039} was first associated with
an EGRET source in \cite{paredes00}, and variability was detected by
\emph{Fermi} between 100 and 300~GeV with a 3.9-d period \citep{abdo09}. The
system is also a TeV emitter, with persistent, variable, and periodic emission,
as detected by H.E.S.S. \citep{aharonian05_ls,aharonian06}. No short-period
pulsations were found that could demonstrate the presence of a pulsar either in
radio \citep{mcswain11} or X-rays \citep{rea11}. We note that traditionally the
name of the optical star, \object{LS~5039}, has also been used to refer to the
binary system, and so we will do in this paper.

In all gamma-ray binaries, the nature of the compact object is fundamental for
understanding the physical processes involved in the particle acceleration that
is responsible for the multi-wavelength emission. If the compact object is a
black hole, the accelerated particles would be powered by accretion, and
produced in the jets of a microquasar (\citealt{paredes06,bosch-ramon06} and the
review in \citealt{bosch-ramon09}). On the other hand, if the compact object is
a young non-accreting pulsar, the particle acceleration would be produced in the
shock between the relativistic wind of the pulsar and the stellar wind of the
massive companion star. This scenario was described in \citet{maraschi81},
\citet{tavani97}, \citet{kirk99}, \citet{dubus06}, \citet{khangulyan07}, and
\citet{bogovalov08, bogovalov12}. Dedicated discussions on the wind-wind
collision scenario for \object{LS~5039} can be found in \citet{khangulyan07},
\citet{sierpowska-bartosik07}, \citet{dubus08}, and \citet{khangulyan08}.
Considerations on the spectral signature of the pulsar wind are provided in
\cite{cerutti08}, on secondary cascading in \citet{bosch-ramon08} and
\citet{cerutti10}, while X-ray absorption and occultation are discussed in
\cite{szostek11}, and limits on the spin-down luminosity of the putative pulsar
in \cite{zabalza11}.

The orbital parameters of \object{LS~5039} such as the eccentricity, masses, and
orbit inclination are not well constrained or are still under discussion
\citep{casares05,aragona09, sarty11}. The proper motion and the space velocity
of the source were first determined in \cite{ribo02}. These authors found that
\object{LS~5039} is escaping from its own regional standard of rest with a total
systemic velocity of 150~km~s$^{-1}$. The past trajectory of \object{LS~5039}
for the last $10^5$~yr computed in \cite{ribo02} marginally suggests an
association with \object{SNR~G016.8$-$01.1}. These authors also discovered an
\ion{H}{i} cavity in the ISM, and argued that it may have been caused by the
stellar wind of \object{LS~5039} or by the progenitor of the compact object in
the system. An in-depth discussion on the supernova process that produced the
compact object in the system \object{LS~5039} can be found in \citet{mcswain02},
\citet{ribo02}, \citet{mcswain04}, and \citet{casares05}. The authors discussed
the expected mass loss during the supernova explosion that produced the compact
object, obtained from the orbital parameters and the mass of the system. This
allowed them to estimate some orbital parameters of the pre-supernova system. We
note that the main limitation for obtaining accurate pre-supernova properties of
\object{LS~5039} is the determination of the orbital parameters, which can be
improved by means of optical observations.

\begin{figure}[] 
\begin{center}
\resizebox{1.0\hsize}{!}{\includegraphics[angle=0]{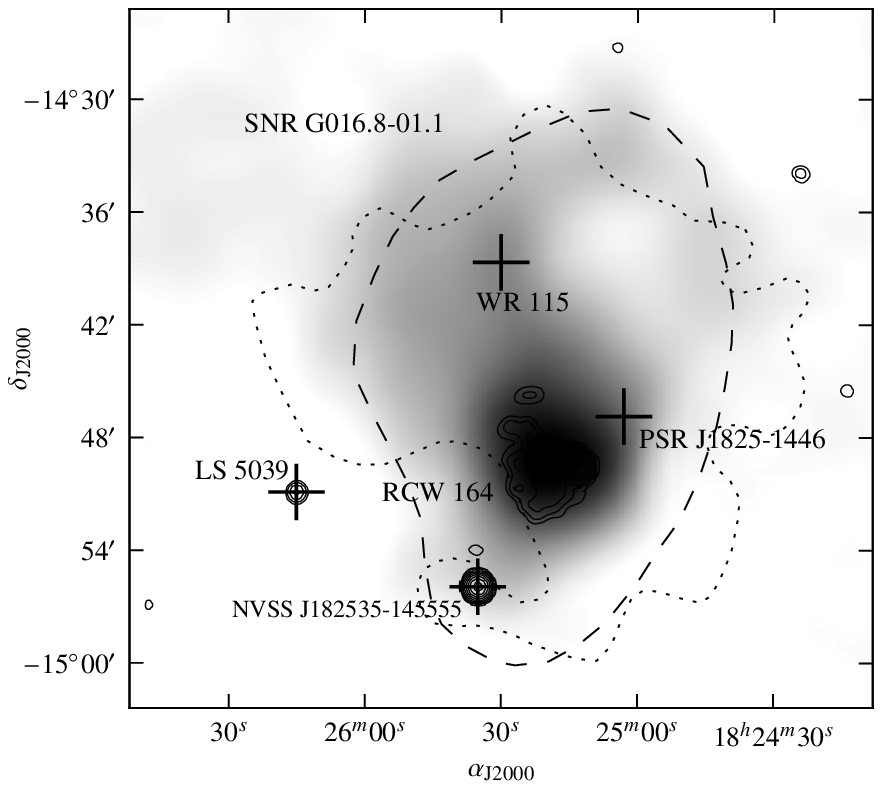}}
\caption{Wide field map around SNR~G016.8$-$01.1. The grey scale corresponds to
the Parkes-MIT-NRAO survey at 5~GHz \citep{tasker94}, and the contours, with
higher resolution, to the NRAO VLA Sky Survey (NVSS) at 1.4~GHz
\citep{condon98}. The dashed line outlines the area covered by the diffuse
emission at 4.75~GHz in \cite{reich86}, and the dotted line the polarised
emission in the same image. These two components trace the morphology of
SNR~G016.8$-$01.1. The more relevant sources in the field are indicated.
\label{fig:intro_wide-field}}
\end{center}
\end{figure}

The region around \object{LS~5039}, shown in Fig.~\ref{fig:intro_wide-field},
contains a complex radio structure that has initially been described in
\cite{reich86}. By means of observations with the Effelsberg 100-m telescope at
1.4, 4.8, and 10~GHz, these authors identified the complex as a composition of
different objects. The brightest source corresponds to the \ion{H}{ii} region
\object{RCW~164}, which coincides with the optical $8^{\prime}\times6^{\prime}$
counterpart in \cite{rodgers60}. It appears to be an optically thick thermal
component that produces the inverted integrated radio emission of the global
structure. Other sources within the structure are a source with 0.11~Jy at
4.8~GHz coincident with the Wolf-Rayet star \object{WR~115} (IC14$-$19 in
\citealt{reich86}), and an unidentified source with 0.13~Jy at 4.8~GHz, also
detected with the VLA as \object{NVSS~J182535$-$145555} \citep{condon98}.
Finally, diffuse emission with a size of $\sim30^{\prime}$ surrounds all the
sources. The source is also visible at 2.7~GHz in the Effelsberg 11~cm radio
continuum survey of the Galactic plane \citep{reich90}. This emission was
identified as a supernova remnant because it is highly polarised (up to 40\% at
4.8~GHz, \citealt{reich86}). The dashed and dotted contours in
Fig.~\ref{fig:intro_wide-field} delimit the diffuse and polarised emission,
respectively, and should trace the shape of the supernova. Unfortunately, the
high contamination of the bright \object{RCW~164} prevents the determination of
the SNR parameters, and consequently its age, distance, centre, and surface
brightness are unknown. The lack of polarization in the direction of
\object{RCW~164} indicates that the SNR is located behind \object{RCW~164},
which is located at $\sim1.8$~kpc \citep{ribo02}. Therefore $\sim2$~kpc is the
lower limit for the SNR distance. The size of the polarised structure is
$\sim30^{\prime}$ and it is approximately centred at $\alpha_{\rm
J2000.0}=18^{\rm h} 25\fm3$ and $\delta_{\rm J2000.0}=-14\degr 46\arcmin$
\citep{reich86,green09}.

\subsection{The nearby isolated pulsar PSR~J1825$-$1446} \label{pulsar}

As part of our project to unveil the origin of the binary \object{LS~5039}, we
searched for other possible sources, such as fossil neutron stars, that may be
related to \object{SNR~G016.8$-$01.1}. The aim was to obtain the proper motion
by means of VLBI of the best candidates to check any possible relation with the
SNR. The most obvious candidate is \object{PSR~J1825$-$1446}, which, in
projection, lies within the SNR itself. To explore other possible candidates we
used the ATNF Pulsar Catalogue,
PSRCAT\footnote{http://www.atnf.csiro.au/research/pulsar/psrcat/}
\citep{manchester05}, which is the most complete and updated pulsar database,
and accepts searches based on pulsar parameters. We restricted the search to
those angular distances for which the pulsars proper motion and age make their
past trajectory compatible with the location of the SNR. We note that this
selection is not complete because, among the known pulsars many do not have good
determinations of the proper motion and/or the age, and their trajectories are
compatible only because of their large uncertainties. We discarded pulsars with
distances below 1.9~kpc and above 9.0~kpc. We found no other candidate with
enough constraints to justify a long-term VLBI project, and therefore we
restricted our observations to \object{PSR~J1825$-$1446}.

\begin{table} 
\setlength{\tabcolsep}{4pt}
\begin{center}
\caption{Parameters of the isolated pulsar PSR~J1825-1446.}
\label{table:psr_properties} 
\begin{tabular}{l c c c}
\midrule
\midrule
Parameter               & Symbol        & Value    & Ref.                          \\
\midrule
Pulsar period           & $P$               & $0.279186875177(5)$~s                 & (1)  \\
Frequency               & $f_{0}$           & $3.58183026823(6)$                    & (1)  \\
Freq. derivative        & $f_{1}$           & $-2.909526(5)\times 10^{-13}$~s$^{-1}$& (1)  \\
Freq. second derivative & $f_{2}$           & $2.24(9)\times 10^{-25}$~s$^{-2}$     & (1)  \\
Characteristic age      & $\tau_{c}$        & $1.95\times 10^5$~yr                  & (1)  \\
Dispersion measure      & DM                & $357\pm5$                             & (1)  \\
Rotation measure        & RM                & $-899\pm10$~rad~m$^{-2}$              & (1)  \\
Distance (DM)           & $d$               & $5.0\pm0.6$~kpc                       & (2)  \\
Spindown luminosity     & $\dot{E}_{\rm sp}$& $4.1\times 10^{34}$~erg~s$^{-1}$      & (1)  \\
Flux density            & $S_{\rm 0.6 GHz}$ & $1.8\pm0.3$~mJy                       & (3)  \\
                        & $S_{\rm 1.1 GHz}$ & $2.4\pm0.5$~mJy                       & (3)  \\
                        & $S_{\rm 1.4 GHz}$ & $3.0\pm0.2$~mJy                       & (4)  \\
                        & $S_{\rm 1.6 GHz}$ & $2.5\pm0.2$~mJy                       & (4)  \\
                        & $S_{\rm 4.8 GHz}$ & $1.2\pm0.1$~mJy                       & (3)  \\
Pulse width (1.4~GHz)   & $W_{\rm 10}$      & 27~ms                                 & (1)  \\
Pulse width (4.8~GHz)   & $W_{\rm 10}$      & 6.8~ms                                & (3)  \\
Polarisation            &                   & $90\pm6\%$                            & (5)  \\
Proper motion           & $\mu_{\alpha}\cos\delta$ & $10\pm18$~mas~yr$^{-1}$        & (1)  \\
                        & $\mu_{\delta}$           & $19\pm115$~mas~yr$^{-1}$       & (1)  \\
\midrule
\end{tabular}
\tablefoot{The proper motion is updated in this paper, see text.}
\tablebib{(1)~\citet{hobbs04}; (2) using the model in \cite{cordes02}; (3) \citet{kijak07}; (4) \citet{lorimer95}; (5) \citet{hoensbroech98}.}
\end{center}
\end{table}

\object{PSR~J1825$-$1446} (B1822$-$14) is an isolated pulsar with a period of
0.28~s discovered in a high radio frequency survey conducted at Jodrell Bank by
\cite{clifton86}. The main properties of the pulsar are described in
Table~\ref{table:psr_properties}. The radio spectrum of
\object{PSR~J1825$-$1446} can be found in \cite{kijak07}, and it presents a
turnover at 1.4~GHz. The spectral index is $-1.4\pm0.7$ above this frequency,
and it is inverted at lower frequencies. The proper motion in
Table~\ref{table:psr_properties} was obtained by means of pulsar timing. The
pulse profile obtained by \cite{gould98} shows a single component at 1.4 and
1.6~GHz. We computed the distance to \object{PSR~J1825$-$1446} using the pulsar
dispersion measure, although it is not a direct measurement because it depends
on the model of the electron distribution in the Galaxy. We used the dispersion
measure quoted in Table~\ref{table:psr_properties} and the electron density
model NE~2001, described in \cite{cordes02}, and obtained a distance of
$5.0\pm0.6$~kpc. The distance uncertainty is 10\%, although it could be
overestimated if the medium along the line of sight of the pulsar is denser than
given by the model.

The possible association between \object{SNR~G016.8$-$01.1} and
\object{PSR~J1825$-$1446} has been discussed in the literature. \cite{clifton86}
already suggested a possible association based only on a positional coincidence.
\cite{clifton92} mentioned the possible relation between them, and the proximity
with the intense \ion{H}{ii} region \object{RCW~164} (see
Fig.~\ref{fig:intro_wide-field}). However, they already found that this
\ion{H}{ii} region appears to be much closer to the Sun than the pulsar, thus
contributing to the DM of the pulsar. The authors also argued that the
probability of having a pulsar in a randomly selected area of the size of
\object{SNR~G016.8$-$01.1} at this Galactic longitude is about 1 in 18, and
therefore the positional proximity may only be by chance. Some notes on
associations between old ($\tau\gtrsim10^{5}$~yr) pulsars and supernova remnants
can be found in \cite{gaensler95}.

\section{Astrometry and proper motion of PSR~J1825$-$1446} \label{obs_psr}

The previously known proper motion of the pulsar, shown in
Table~\ref{table:psr_properties}, has large uncertainties and did not allow us
to study the origin of this pulsar. The astrometry in the literature, detailed
in Sect.~\ref{archival_psr}, is scarce. Therefore, we conducted a dedicated 2-yr
VLBI astrometric project to measure the proper motion of the pulsar.

\subsection{Radio VLBI observations} \label{obs_psr_radio}

Before starting the VLBI project, we conducted a short VLA observation at
8.6~GHz to obtain the initial position required for the correlation of the VLBA
data. The VLA observation was conducted on November 14, 2008, in the A
configuration of the VLA, and lasted 30 minutes. At that epoch the VLA was
performing the upgrade to the EVLA. Seventeen antennas used the new EVLA feeds,
eight used the old VLA feeds, and four antennas were not available. We observed
the source during three scans of 3.6~minutes, bracketed by 1.6-minute scans on
the phase calibrator J1832$-$1035 (1832$-$105). The correlation position of the
reference source J1832$-$1035 was $\alpha_{\rm J2000.0}=18^{\rm h} 32^{\rm m}
20\fs8436$ and $\delta_{\rm J2000.0}=-10\degr 35\arcmin 11\farcs299$. Standard
data reduction in AIPS\footnote{The NRAO Astronomical Image Processing System.
http://www.aips.nrao.edu/} was performed, including standard normalization of
the flux density between EVLA and non-EVLA antennas. The data were imaged with
an intermediate robust parameter of 0 within AIPS, as a compromise between
angular resolution and sensitivity, providing a synthesized beam of
$460\times210$~mas at a position angle (P.A.) of 37$^{\circ}$. We used a cell
size of 10~mas. The image shows a point-like source with a peak flux density of
$0.76\pm0.07$~mJy~beam$^{-1}$, and a total flux density of $0.8\pm0.1$~mJy. The
pulsar position was measured with the AIPS task JMFIT.

The VLBI campaign consisted of three observations with the Very Long Baseline
Array (VLBA) at 5~GHz separated by one year. The observations were conducted on
the same day (May 4) in 2009, 2010, and 2011, during six hours between UTC~07:40
and 13:40. Observations separated by one year prevent any annual parallax
displacement, which is expected to have a maximum amplitude of $\sim0.2$~mas for
a source at 5~kpc. Additionally, identical schedules were chosen to minimise any
observational difference between epochs (e.g. source elevation, scan structure).
The observations were conducted at a frequency of 5.0~GHz (6~cm wavelength),
which was chosen as a compromise between angular resolution, spectral index of
the pulsar ($\alpha\sim-1.4$), and the quality of the best VLBI calibrator next
to the source, \object{J1825$-$1718}, which is partially resolved at long
baselines due to interstellar scattering. The first epoch was recorded with a
total data rate of 256~Mbps per station distributed in four dual polarisation
subbands, each of them with a bandwidth of 8~MHz, and recorded using
16~frequency channels, two-bit sampling, and 2~s of integration time. The
correlation of this first epoch was performed in the hardware VLBA correlator at
Socorro. The second and third epochs, observed and correlated in 2010 and 2011,
respectively, were recorded at the available bit rate of 512~MHz, and eight
subbands were used, which improved the sensitivity of the observation. The 2010
and 2011 correlations were performed with the DiFX \citep{deller07} software
correlator at Socorro. The data from the three observations of
\object{PSR~J1825$-$1446} were also correlated using pulsar gating (see
Appendix~\ref{obs_psr_gating}).

The observations were performed using phase-referencing on the nearby phase
calibrator \object{J1825$-$1718}, located at 2.5$^{\circ}$ from the pulsar,
switching between them with a cycling time of four minutes, which is compatible
with the expected coherence time. The phase calibrator was correlated at
$\alpha_{\rm J2000.0}=18^{\rm h} 25^{\rm m} 36\fs53228$ and $\delta_{\rm
J2000.0}=-17\degr 18\arcmin 49\farcs8485$. However, for the final fit result we
used the last available position of the source in the frame of ICRF (see
Sect.~\ref{obs_psr_astrometry}). As an astrometric check source we observed the
quasar J1844$-$1324, which is separated 5$^{\circ}$ from the pulsar and
6$^{\circ}$ from the phase calibrator. The source has a flux density of 65~mJy,
which, in natural weighted images, is dominated by a compact core that accounts
for 94\% of the total emission, and displays slightly extended emission
eastwards. Only the compact core is seen in the uniformly weighted images used
to measure its position. Two scans of the fringe finder J1733$-$1304 were
observed at each epoch.

\begin{table} 
\begin{center}
\caption{Ionospheric correction applied to the VLBA data on PSR~J1825$-$1446. Units are mas.}
\label{table:ionosphere} 
\begin{tabular}{crrcrrc}
\midrule
\midrule
Epoch   & \multicolumn{2}{c}{Mean offset}&  & \multicolumn{2}{c}{Dispersion}& Best   \\
 \cmidrule{2-3}     \cmidrule{5-6}  
        & \multicolumn{1}{c}{$\alpha$}      & \multicolumn{1}{c}{$\delta$}           &  & \multicolumn{1}{c}{$\alpha$}          & \multicolumn{1}{c}{$\delta$}                &        \\
\midrule
2009    & 0.13     & 0.92          &  & 0.04           & 0.17              & CODE    \\
2010    & $-$0.05  & 1.96          &  & 0.09           & 0.36              & IGS    \\
2011    & 0.03     & 2.73          &  & 0.19           & 0.35              & IGS    \\
\midrule
\end{tabular}
\end{center}
\tablefoot{CODE: Center for Orbit Determination in Europe. IGS: the International GNSS Service}
\end{table}

The data reduction was performed in AIPS. Flagging based on predicted off-source
times, owing to slewing or failures, was applied using UVFLG. A priori
visibility amplitude calibration used the antenna gains and the system
temperatures measured at each station. We used ionospheric total electron
content (TEC) models based on GPS data obtained from the CDDIS data
archive\footnote{The Crustal Dynamics Data Information System
http://cddis.nasa.gov/} to correct the data from the sparse ionospheric
variations. Several ionospheric models are produced each day by different groups
(i.e. the Jet Propulsion Laboratory (JPL), the Center for Orbit Determination in
Europe (CODE), the Geodetic Survey Division of Natural resources Canada (EMR),
the ESOC Ionosphere Monitoring Facility (ESA), and the Universitat
Polit\`{e}cnica de Catalunya (UPC), among others). We applied the corrections
from all available models to the data to check the consistency between models.
There are currently 12 models available for the data obtained in 2009, 16 for
2010, and 19 for 2011. In Table~\ref{table:ionosphere} we show the offset
position between the uncorrected data and the corrected data (average value
considering all models) for right ascension and declination, respectively. We
also show the standard deviation of the positions measured with the different
models, and the model that provides a position closer to the mean value, which
was used for the final data reduction.

Standard instrumental corrections were applied (parallactic angle, instrumental
offsets, and bandpass corrections). Fringe-fitting on the phase calibrator was
performed with the AIPS task FRING. The amplitude and phase calibration, flags,
and bandpass correction tables were applied to the target pulsar and the
astrometric check source data, which were averaged in frequency, and clean
images were produced with IMAGR. A cell size of 0.1~mas was used for cleaning
all images, which were produced using a weighting scheme with robust parameter
$-$2 within AIPS. A tapering of 60~M$\lambda$ was applied to avoid the use of
phases where the correlated flux density of the calibrator was below
$\sim$40--10~mJy. No self-calibration of the data was possible because of the
low flux density of the target source.

\begin{table} 
\begin{center}
\caption{Parameters of the observations of PSR~J1825$-$1446, and the resulting
images. The last column is the improvement factor of the signal-to-noise ratio
achieved when using pulsar gating.}
\label{table:psr_obs} 
\begin{tabular*}{\columnwidth}{@{\extracolsep{-3.5pt}}c c c r@{ at~~~}r c c c @{}}
\midrule
\midrule
MJD      & Array & Freq.    & \multicolumn{2}{c}{HPBW}                  & $(S/N)_{\rm 0}$  & $(S/N)_{\rm G}$  &  Gain  \\
         &       & [GHz]    & \multicolumn{2}{c}{[mas$^{2}$ at $^{\circ}$]}&           &          &         \\
\midrule
55607.83 & VLA   & 8.6      & 460$\times$210        & 37                & 11           & ---      & ---     \\
\midrule
54955.44 & VLBA  & 5.0      & 4.9$\times$2.2        & 5                 & 3.5          & 9.8      & 2.8     \\
55320.44 & VLBA  & 5.0      & 6.2$\times$2.7        & 6                 & 6.6          & 11.5     & 1.7     \\
55685.44 & VLBA  & 5.0      & 6.9$\times$2.9        & 1                 & 6.1          & 8.4      & 1.4     \\
\midrule
\end{tabular*}
\end{center}
\end{table}

The source has a declination of $-15^{\circ}$, and the maximum elevation for
most of the VLBA antennas is $40^{\circ}$. Some phase decoherence was present
when using the data from the beginning and the end of the observations,
presumably as a consequence of the atmospheric effects caused by the low
elevation of the source and the phase calibrator. To reduce the astrometric
errors, we flagged all visibilities with elevations below $20^{\circ}$, and only
used the data closer to the culmination of the source, discarding about
1.5~hours at the beginning and at the end of the observations. To reduce the
effects of the extended emission of the astrometric check source, visibilities
with $uv$-distances below 10~M$\lambda$ were not used. The positions of
\object{PSR~J1825$-$1446} and the astrometric check source were measured by
fitting a Gaussian component with JMFIT within AIPS. In
Table~\ref{table:psr_obs} we show some parameters of the observation, the
resolution, and a comparison of the obtained signal-to-noise ratio with and
without pulsar gating (see Appendix~\ref{obs_psr_gating} for the details). The
position errors of \object{PSR~J1825$-$1446} were finally computed as the
standard deviation of the three positions of the astrometric check source. The
final positions measured in the images from the data correlated with pulsar
gating are shown in Table~\ref{table:psr_positions}.

\begin{table*} 
\begin{center}
\caption{Astrometry and fit residuals of PSR~J1825$-$1446.}
\label{table:psr_positions} 
\begin{tabular}{cc cc lc lc rr}
\midrule
\midrule
MJD   & Epoch   & Array & Freq. &  $\alpha_\mathrm{J2000}$   & $\sigma_{\alpha\cos\delta}$ &  $\delta_\mathrm{J2000}$ & \multicolumn{1}{c}{$\sigma_{\delta}$} &  \multicolumn{1}{c}{Res.$_{\alpha\cos\delta}$} & \multicolumn{1}{c}{Res.$_{\delta}$}  \\
      & Y-M-D   &       &  [GHz]& [h, m,  s]        & \multicolumn{1}{c}{[mas]}   & [$^{\circ},~^{\prime},~^{\prime\prime}$]       & \multicolumn{1}{c}{[mas]}             & \multicolumn{1}{c}{[mas]}               & \multicolumn{1}{c}{[mas]}                   \\
\midrule
49060.53 & 1993-03-14   & VLA-B   & 8.4 & 18~~25~~2.94(3)    & 40                     & $-$14~~46~~52.77(6)   &  60                  &  $-$37   &   $-$61  \\
49480.00 & 1994-05-08   & Timing  & 0.4--1.6  & 18~~25~~2.927(5)   & 75               & $-$14~~46~~52.6(5)    & 500                  & $-$245   &     142 \\
54784.04 & 2008-11-14   & VLA-A   & 8.6 & 18~~25~~2.9539(8)  & 12                     & $-$14~~46~~53.163(15) &  15                  &     13.4 &  $-$0.0  \\
54955.44 & 2009-05-04   & VLBA    & 5.0 & 18~~25~~2.95324(4) & 0.6\tablefootmark{a}   & $-$14~~46~~53.1767(5) & 0.5\tablefootmark{b} &   $-$0.6 &  $-$0.3  \\
55320.44 & 2010-05-04   & VLBA    & 5.0 & 18~~25~~2.95398(4) & 0.6\tablefootmark{a}   & $-$14~~46~~53.2047(5) & 0.5\tablefootmark{b} &      0.7 &     0.7  \\
55685.44 & 2011-05-04   & VLBA    & 5.0 & 18~~25~~2.95460(4) & 0.6\tablefootmark{a}   & $-$14~~46~~53.2347(5) & 0.5\tablefootmark{b} &      0.0 &  $-$0.4  \\
\midrule
\end{tabular}
\tablefoot{
\tablefoottext{a}{An additional uncertainty of 0.9~mas in right ascension due to the global astrometry errors on the phase-reference source position should be included when considering this position in the ICRF.}
\tablefoottext{b}{The phase-reference source has an additional uncertainty of 1.4~mas in declination in the ICRF.}
}
\end{center}
\end{table*}

\subsection{Archival astrometry of PSR~J1825$-$1446} \label{archival_psr}

To measure the proper motion of \object{PSR~J1825$-$1446} we searched all
currently available astrometry of this isolated pulsar. \cite{frail97} imaged
the field around several pulsars with the VLA. They observed
\object{PSR~J1825$-$1446} (B1822$-$14) on March 14, 1993 at 8.4~GHz for two
scans of 10~minutes each. The observations were conducted with the VLA in its B
configuration. The phase calibrator was J1832$-$1035 (J1832$-$105, 1829$-$106).
We have re-analysed the data to measure the position of the pulsar. Although the
scan duration is longer than optimal for astrometry, the pulsar is detected as a
point-like source with a peak flux density of $0.78\pm0.10$~mJy~beam$^{-1}$ when
using uniform weighting, which yields a synthesized beam size of
$1.0\times0.6$~arcsec$^{2}$. We transformed the positions of the observation
from B1950 to J2000, and we updated the position of the phase calibrator to the
best currently known position, quoted in Sect.~\ref{obs_psr_radio}. We fitted
the position using JMFIT within AIPS, and the errors correspond to formal errors
of a Gaussian fit. The fitted position can be found in
Table~\ref{table:psr_positions}.

\begin{figure*}[] 
\begin{center}
\resizebox{1.0\hsize}{!}{\includegraphics[angle=0]{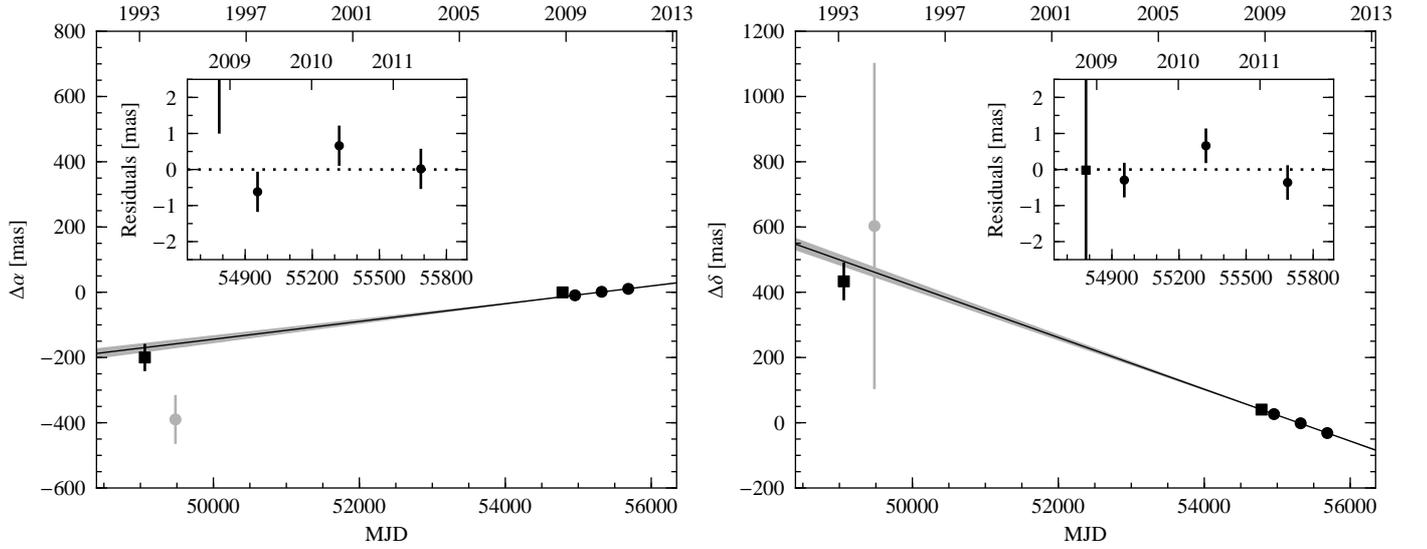}}
\caption{Position of PSR~J1825$-$1446 with respect to time (MJD in lower axis
and year in top axis) in right ascension (\emph{left}) and declination
(\emph{right}). The solid black lines represent the fitted proper motion, and
the grey area the uncertainty of the fit at 3-$\sigma$ level. The black squares
represent the VLA positions, the black circles indicate the VLBA positions, and
the grey circle the position from pulsar timing (not used in the fit). The inset
shows the position residuals from the fitted model for the 2008--2012 period.
The source positions are plotted with uncertainties at 1-$\sigma$ level.
\label{fig:pm_psr}}
\end{center}
\end{figure*}

The position of \object{PSR~J1825$-$1446} found in \cite{hobbs04}, obtained by
means of pulsar timing, is also shown in Table~\ref{table:psr_positions}. The
position is in the solar system centre of mass, or barycentric, frame, and it
was obtained by fitting the pulsar period and period derivative, the ecliptic
position and proper motion, among other variables, to the time of arrival (TOA)
of the pulses during several years. However, we can see in
Table~\ref{table:psr_positions} that the timing position differs from the
expected position by 245~mas (3-$\sigma$) in right ascension, which appears to
indicate that the uncertainty may be underestimated (see also
Fig.~\ref{fig:pm_psr} and the difference with the declination error). The
difference between the errors in right ascension and declination can be
understood considering the pulsar position. The pulsar is located near the
ecliptic plane (with ecliptic latitude $\beta=8\fdg53$), which has two
consequences in the errors of the ecliptic coordinates ($\lambda,\beta$). In
particular, the uncertainty in $\lambda$ and $\beta$ are inversely proportional
to $\cos\beta$, and $\sin\beta$,
respectively\footnote{http://www.cv.nrao.edu/course/astr534/PulsarTiming.html}.
For \object{PSR~J1825$-$1446} we have that $(\cos\beta)^{-1}\sim1.0$, and
$(\sin\beta)^{-1}\sim6.7$, which intrinsically yields larger errors in the
determination of the ecliptic latitude by means of pulsar timing. Large errors
in the ecliptic latitude $\beta$ produce the large error in declination, which
is indeed a factor 6.7 larger than the uncertainty in right ascension. Another
possible source of error is the pulsar timing fit solution. \cite{hobbs04}
fitted 197 time of arrival (TOA) data points spanning 17.4~years and obtained a
proper motion of $\mu_{\alpha}\cos\delta=10\pm18$ and
$\mu_{\delta}=19\pm115$~{mas~yr}$^{-1}$. The fitting time residuals were
7.539~ms. If this value were understimated, this would explain the small error
in right ascension. Another possible source of error is the uncertainty in the
link between the barycentric celestial reference system (BCRS) and the ICRF.
However, it should be well below the 245~mas difference found between the timing
position and the expected position at that epoch; for example
\cite{chatterjee09} found differences between both reference frames below 1~mas.
We have not found any clear explanation to justify the discrepancy between the
right ascension position given by pulsar timing and the expected one. Given the
large difference, and the intrinsic difference between an interferometric
measurement and a timing fit, we have not included the timing position in our
final determination of the proper motion of \object{PSR~J1825$-$1446}, although
we have included the position in Table~\ref{table:psr_positions} and
Fig.~\ref{fig:pm_psr} for reference.

\subsection{Proper motion of PSR~J1825$-$1446} \label{obs_psr_astrometry}

To obtain the proper motion, we used the two VLA and the three VLBA observations
presented in Sect.~\ref{obs_psr}. The reference position of the phase calibrator
for the VLA observations was updated to the currently known position of the
phase calibrator, quoted in Sect.~\ref{obs_psr_radio}. All VLBA positions were
updated according to the most recent available position\footnote{GSFC 2010a
astro solution, from sched v10.0} of \object{J1825$-$1718} $\alpha_{\rm
J2000.0}=18^{\rm h} 25^{\rm m} 36\fs532278\pm0\fs00006$ (or $\pm0.9$~mas) and
$\delta_{\rm J2000.0}=-17\degr 18\arcmin 49\farcs8482\pm0\farcs0014$ (or
$\pm1.4$~mas). This global uncertainty of the reference source position should
be included when working in the ICRF. However, in this case it is only a common
offset shared by all VLBA observations. We have computed the proper motion fit
using the position of \object{J1825$-$1718} as a reference. This means that the
0.9 and 1.4~mas uncertainties have been included in all the non-VLBA (but not in
the VLBA ones), to compute the proper motion, and it has been added afterwards
to the final fitted position. All positions and uncertainties are shown in
Table~\ref{table:psr_positions}. We fitted the proper motion of the source as a
linear model of position as a function of time for right ascension and
declination separately. We fitted the data using a Levenberg-Marquardt nonlinear
least-squares fit \citep{marquardt63}. The solution is highly dominated by the
three VLBA observations, which were performed in the same scheduled conditions
(see Sect.~\ref{obs_psr_radio}). The reference epoch is computed as the weighted
mean MJD, which is MJD~55291.525. The obtained astrometric parameters for
\object{PSR~J1825$-$1446} are

\begin{align*}
\alpha_\mathrm{J2000}   & =  18^{\rm h}\, 25^{\rm m}\, 2\fs95389\, \left(\pm0.3\pm0.9~\mathrm{mas}\right),     \\
\delta_\mathrm{J2000}   & = -14^{\circ}\, 46^{\prime}\, 53\farcs2031\, \left(\pm0.3\pm1.4~\mathrm{mas}\right),  \\
\mu_\alpha \cos{\delta} & =  10.0 \pm 0.3\ {\rm mas\ yr}^{-1},  \\
\mu_\delta              & = -29.0 \pm 0.3\ {\rm mas\ yr}^{-1},  \\
\end{align*}

\noindent
where the first set of errors in the positions corresponds to statistical
uncertainties, and the second set corresponds to the uncertainty in the phase
reference position. The degrees of freedom of the fit are 4, and the reduced
$\chi^{2}$ is 0.95 and 0.98 in right ascension and declination, respectively.

\section{Astrometry and proper motion of LS~5039} \label{obs_ls5039}

To fit the proper motion of \object{LS~5039} we used new interferometric radio
data together with archival radio and optical data. In this section we will only
describe in detail the data reduction of three radio projects. The description
of the other data used is already published, or will be published elsewhere. In
particular we will describe a VLA+PT observation conducted in 2003, three VLBA
observations conducted between 2004 and 2006, and a short EVN observation
conducted in 2011. The parameters of these observations are summarised in
Table~\ref{table:ls_vlbaobs}. 

\begin{table} 
\begin{center}
\caption{Parameters of the LS~5039 observations described in Sect.~\ref{obs_ls5039}, parameters of the resulting images, and the total and peak flux densities measured.}
\label{table:ls_vlbaobs}
\begin{tabular*}{\columnwidth}{@{}c@{\hspace{5pt}} c@{\hspace{6pt}} c@{\hspace{6pt}} r@{ at~~}r @{\hspace{6pt}} r@{ $\pm$ }l r@{ $\pm$ }l@{}}
\midrule
\midrule
MJD      & Array & Freq.    & \multicolumn{2}{c}{HPBW}                  & \multicolumn{2}{c}{$S_{\nu}$}  & \multicolumn{2}{c}{$S_{\rm peak}$}   \\
         &       & [GHz]    & \multicolumn{2}{c}{[mas$^{2}$ at $^{\circ}$]} & \multicolumn{2}{c}{[mJy]}  & \multicolumn{2}{c}{[mJy~b$^{-1}$]}          \\
\midrule
52856.20 & VLA+PT & 15.0      & 148$\times$39         &  23             &  9.6 & 0.5                     & 9.0 & 0.3     \\
53167.35 & VLBA   &  8.5      & 4.5$\times$2.2        &   7             & 10.9 & 1.5                     & 6.0 & 0.6     \\
53532.35 & VLBA   &  8.4      & 2.8$\times$1.1        &   6             &  7.1 & 0.9                     & 2.6 & 0.3     \\
53764.71 & VLBA   &  8.4      & 3.2$\times$1.0        &  10             &  4.4 & 0.7                     & 3.1 & 0.3     \\
55635.15 & EVN    &  8.4      & 5.6$\times$1.3        &  $-$15          &  8.0 & 2.0                     & 6.1 & 1.0     \\
\midrule
\end{tabular*}
\end{center}
\end{table}

\subsection{VLA-PT observation} \label{obs_ls5039_vlapt}

We observed \object{LS~5039} with the VLA plus the VLBA antenna in Pie Town (PT)
at 15.0~GHz on August 05, 2003. We used the VLA in its A configuration plus PT,
which is located at a distance of 52~km from the VLA site, and provides longer
baselines with all the VLA antennas. The observation spanned from 01:00 to
08:40~UTC, and was centred at MJD~52856.2. The phase reference calibrator was
J1832$-$1035 (1832$-$105), located at 4.5$^{\circ}$ from \object{LS~5039}. Two
IFs of 50~MHz were used. The cycle time including the phase calibrator and the
target source was 3~minutes. The ICRF source \object{J1911$-$2006} (1911$-$201)
was observed regularly during the observations. The flux calibrator was
\object{3C~286} (1331+305). The orbital phase of the binary system at the time
of the observation was 0.92, using the ephemeris from \cite{casares12_ls}.

Standard reduction procedures were applied to the data, and a precise position
of the source was obtained when using a uniform weighting scheme (robust $-$5
within AIPS), and a cellsize parameter of 5~mas. The source position, fitted
with JMFIT within AIPS, has an uncertainty of 1 and 2~mas in right ascension and
declination, respectively. The observational details, the obtained synthesized
beam size, and the fitted components are shown in Table~\ref{table:ls_vlbaobs}.

\subsection{VLBA astrometric project} \label{obs_ls5039_vlba}

The VLBA projects BD087 and BD105 are part of a long-term astrometric project to
measure the proper motion and parallax of several X-ray binaries. The segment
BD087G, and the continuations BD105A and BD105G observed \object{LS~5039} at a
frequency of $\sim$8.5~GHz. The observations took place on June 11, 2004, June 11,
2005, and January 29, 2006, and lasted 6, 6, and 5~hours, respectively. The
three observations were recorded with a total data bit rate of 128~Mbps. In the
first epoch, the data rate was distributed in 8~IF channels with a bandwidth of
4~MHz each, with one single polarisation (right circular polarisation) and two-bit
sampling, while the latter two used two IF channels with 8~MHz bandwidth and dual
circular polarisation, also recorded with two-bit sampling.

The phase reference source was \object{J1825$-$1718}, which is slightly resolved
also at 8.4~GHz (although it is much more compact than at 5~GHz), and is located
at $2.47^{\circ}$ from \object{LS~5039}. The cycling time of the
phase-referenced observations was three minutes. Several check sources were
observed during the three epochs. In particular, the sources
\object{N1834$-$1433}, \object{N1819$-$1419}, and \object{N1818$-$1108} were
observed in all epochs and were used to estimate the astrometric uncertainties.
The fringe finder was \object{J1733$-$1304}. The data reduction is analogous to
the one described in Sect.~\ref{obs_psr_radio}. In this case, an statistical
approach to determine the best ionospheric correction was not applied, because
the effect is smaller at this frequency. We note that the correction of the
Earth orientation parameters was not applied to the data because it did not
affect the position measurement, and therefore we preferred not to apply this
additional correction to the datasets.

Owing to the lower flux density of \object{LS~5039} at 8.4~GHz, the three epochs
were imaged using a weighting scheme with robust parameter 0 within AIPS, as a
compromise between spatial resolution and signal-to-noise ratio. To reduce the
effect of the source extended emission (see Sect.~\ref{uncertainties_ls}) we
omitted the baselines with shortest $uv$-distances. We found a compromise
between significance of the detection, and the $uv$-distance reduction by
cutting the baselines below 20, 30, and 40~M$\lambda$ for the three epochs. For
the first epoch a lower threshold was required because the data quality was
worse. On the other hand, the longest baselines had to be softened because of
the poor quality of the partially resolved phase reference source. We reduced
the weight of baselines longer than a certain value. For the first epoch we used
a tapering parameter of 60~M$\lambda$. For the latter two projects, the tapering
was at 120~M$\lambda$. The resulting resolution and fitted components are shown
in Table~\ref{table:ls_vlbaobs}.

\subsection{EVN observations} \label{obs_ls5039_evn}

The most recent VLBI observation of \object{LS~5039} was performed with the
European VLBI Network (EVN). We observed several sources around \object{LS~5039}
to search for possible phase calibrators and in-beam calibrators, and test the
longest EVN baselines to check the viability of an astrometric project. Among
other sources, we observed the phase calibrator \object{J1825$-$1718}, also used
in the previously described VLBI observations. The observations were conducted
on March 15, 2011, between UTC~02:00 and 05:00. We only observed 11 scans on
\object{LS~5039}, bracketed by scans on the phase calibrator with a cycling time
of 3.4~minutes and distributed along the observation. The total time on-source
was approximately 20~minutes. The antennas included in the observation were
Effelsberg, Westerbork, Onsala, Medicina, Yebes, Svetloe, Badary,
Zelenchukskaya, Nanshan (Urumqi), Sheshan (Shanghai), and Hartebeesthoek. The
data were recorded with a total data bit rate of 1024~Mbps, providing a
high-sensitivity array. The correlated parameters are eight IF channels with
both circular polarisations, each of them with 16~MHz bandwidth (provided by
32~frequency channels of 500~kHz), and two-bit sampling.

The data reduction is analogous to the one described in
Sect.~\ref{obs_psr_radio} and Sect.~\ref{obs_ls5039_vlba}. No useful FRING
solutions were found between Hartebeesthoek and the rest of the antennas, which
considerably reduced the North-South array extend, and therefore the resolution
in declination on the final image. After the phase calibration from
\object{J1825$-$1718}, the North-South maximum baseline was of about
20--40~M$\lambda$, while in the East-West direction it was of
160--180M$\lambda$, although large gaps were present in the $uv$-plane due to
the short observation time. We note that large phase instabilities appeared due
to the low declination of the source. We finally imaged the source with robust
parameter 0, minimum $uv$-distance set to 15~M$\lambda$, and $uv$-taper set to
120~M$\lambda$. Even with the long maximum baselines, the final synthesized beam
size is 5.6$\times$1.3~mas$^{2}$ in P.A. of $-15^{\circ}$. The resolution of the
images and the fitted components are shown in Table~\ref{table:ls_vlbaobs}.

\subsection{Archival data of LS~5039} \label{archival_ls5039}

Astrometry of \object{LS~5039} can be obtained from two types of observations.
On one hand we have positions from optical observations of the main star of the
system. This observations, coming from large (and global) astrometric projects,
cover a long time, of the order of decades, although they provide limited
astrometric accuracy of about 10--300~mas. On the other hand, radio
interferometric observations provide observations of the radio nebula around the
binary system. These observations have been obtained in the last $\sim$10~years,
but their uncertainties are between 1 and 10~mas. The time span of the optical
observations can be combined with the precise radio positions to fit an accurate
position and proper motion of the source. However, combining datasets with
considerable different properties brings important caveats that have to be taken
into account. These will be described in Sect.~\ref{movprop_ls5039}. The
astrometry described here is summarized in Table.~\ref{table:ls_positions}. The
corresponding orbital phases were computed using the ephemerides in
\cite{casares12_ls}.

\begin{table*} 
\begin{center}
\caption{Astrometry and fit residuals of LS~5039 for optical and radio projects.}
\label{table:ls_positions} 
\begin{tabular*}{\textwidth}{@{\extracolsep{-0pt}}lccrlrlrrr}
\midrule
\midrule
Project &  MJD   & Epoch   & \multicolumn{1}{c}{$\phi_{\rm orb}$} & $\alpha_\mathrm{J2000}$ & $\sigma_{\alpha\cos\delta}$ &  $\delta_\mathrm{J2000}$                & \multicolumn{1}{c}{$\sigma_{\delta}$} &  \multicolumn{1}{c}{Res.$_{\alpha\cos\delta}$} & \multicolumn{1}{c}{Res.$_{\delta}$}   \\
        &        &  Y-M-D  &                  & [h, m,  s]              & \multicolumn{1}{c}{[mas]}   & [$^{\circ},~^{\prime},~^{\prime\prime}$]& \multicolumn{1}{c}{[mas]}             & \multicolumn{1}{c}{[mas]}                      & \multicolumn{1}{c}{[mas]}                    \\
\midrule
AC2000.2  (O)  & 17408.47 & 1906-07-17 & 0.8(2)  & 18~~26~~15.015(8)     & 120                  &  $-$14~~50~~53.27(11)    & 110                    &     40  &  160      \\
USNO-A2.0  (O) & 33857.61 & 1951-07-30 & 0.0(1)  & 18~~26~~15.034(17)    & 260                  &  $-$14~~50~~53.59(25)    & 250                    &     10  &  230      \\
TAC2.0  (O)    & 44050.66 & 1979-06-26 & 0.57(5) & 18~~26~~15.056(5)     &  80                  &  $-$14~~50~~54.08(5)     &  50                    &    140  &  $-$8     \\
GSC1.2  (O)    & 46669.35 & 1986-08-27 & 0.99(4) & 18~~26~~15.06(2)      & 300                  &  $-$14~~50~~54.3(3)      & 300                    &     80  & $-$145    \\
Tycho-2  (O)   & 48559.95 & 1991-10-30 & 0.01(3) & 18~~26~~15.043(10)    & 150                  &  $-$14~~50~~54.23(12)    & 120                    & $-$150  &  $-$55    \\
AP357  (R)     & 50913.00 & 1998-04-10 & 0.43(2) & 18~~26~~15.0559(7)    &  10                  &  $-$14~~50~~54.240(10)   &  10                    &    5.5  & $-$8.1    \\
2MASS  (O)     & 51401.02 & 1999-08-11 & 0.37(2) & 18~~26~~15.056(4)     &  60                  &  $-$14~~50~~54.23(6)     &  60                    & $-$7.3  &   18      \\
UCAC1  (O)     & 51649.00 & 2000-04-15 & 0.85(2) & 18~~26~~15.0563(9)    &  13                  &  $-$14~~50~~54.277(13)   &  13                    & $-$2.9  &$-$28      \\
GR021A  (R)    & 51698.40 & 2000-06-03 & 0.49(2) & 18~~26~~15.05645(10)  & 1.5\tablefootmark{a} &  $-$14~~50~~54.2512(15)  &   2\tablefootmark{b}   & $-$1.6  & $-$0.5    \\
AP453  (R)     & 52856.20 & 2003-08-05 & 0.92(2) & 18~~26~~15.05812(11)  & 1.6~~                &  $-$14~~50~~54.280(2)    &   2~~                   &    1.0  & $-$1.6    \\
BD087G  (R)    & 53167.35 & 2004-06-11 & 0.58(2) & 18~~26~~15.05831(14)  & 2\tablefootmark{a}   &  $-$14~~50~~54.2865(20)  &   2\tablefootmark{b}   & $-$2.3  & $-$0.7    \\
BD105A  (R)    & 53532.35 & 2005-06-11 & 0.02(2) & 18~~26~~15.05890(8)   & 1.3\tablefootmark{a} &  $-$14~~50~~54.2964(14)  &   1.4\tablefootmark{b} & $-$0.4  & $-$1.7    \\
BD105G  (R)    & 53764.71 & 2006-01-29 & 0.51(2) & 18~~26~~15.05916(7)   & 1.1\tablefootmark{a} &  $-$14~~50~~54.2996(12)  &   1.2\tablefootmark{b} & $-$1.1  &    0.6    \\
EF018A--C (R)  & 54160.71 & 2007-03-01 & 0.89(2) & 18~~26~~15.0597(2)    & 4\tablefootmark{a}   &  $-$14~~50~~54.3092(12)  &   1.2\tablefootmark{b} & $-$0.2  &    0.5    \\
BR127A--E (R)  & 54290.27 & 2007-07-09 & 0.06(2) & 18~~26~~15.06003(8)   & 1.3\tablefootmark{a} &  $-$14~~50~~54.3096(12)  &   1.2\tablefootmark{b} &    1.8  &    3.3    \\
EM085  (R)     & 55635.15 & 2011-03-15 & 0.37(2) & 18~~26~~15.06184(8)   & 1.3\tablefootmark{a} &  $-$14~~50~~54.3452(20)  &   2.0\tablefootmark{b} &    2.7  & $-$0.2    \\
BC196  (R)     & 55803.13 & 2011-08-30 & 0.37(2) & 18~~26~~15.06188(7)   & 1.1\tablefootmark{a} &  $-$14~~50~~54.3511(14)  &   1.4\tablefootmark{b} &    1.4  & $-$2.0    \\
\midrule
\end{tabular*}
\tablefoot{
\tablefoottext{a}{An additional uncertainty of 0.9~mas in right ascension due to the global astrometry errors on the phase-reference source position should be included when considering this position in the ICRF.}
\tablefoottext{b}{The phase-reference source has an additional uncertainty of 1.4~mas in declination in the ICRF.}
}
\end{center}
\end{table*}

\subsubsection{Archival optical astrometry of LS~5039} \label{optical_ls5039}

Optical observations of \object{LS~5039} are available from historical
photographic catalogues since 1905. Updated global solutions have been obtained
for the historical photographic plates by using more recent reference
catalogues. The updated versions of the classical catalogues provide improved
systematic uncertainties on the star position. However, some newer versions
combine catalogues from different epochs to fit a global solution that considers
the proper motion of the stars. Therefore, they are not suitable for our proper
motion determination because they are not independent, and they include a proper
motion estimation in their own position fitting. This is the case, for example,
of the Palomar Observatory Sky Survey plates used to produce the USNO-A2.0
catalogue \citep{monet98}, which uses only the blue and red original plates,
while the new version of the catalogue, USNO-B1.0 \citep{monet03}, with better
uncertainties, uses data from 1949 to 2002, combined using a proper motion of 4,
and $-14$~mas~yr$^{-1}$ in right ascension and declination, respectively. It is
also the case of GSC 1.2 compared to the newer versions (GSC~2.2 and GSC~2.3.2),
or UCAC1 when compared to UCAC2 and UCAC3, which have improved references and
systematic error treatment. Therefore, although new catalogues are currently
available, we have used only the versions that are not compilation catalogues.
This means that most of the optical values used here were already available at
the time of writing of \cite{ribo02}. 

In particular, for the proper motion determination we used the following
catalogues. AC2000.2 is the second version of the Astrographic Catalog 2000 and
provides accurate optical positions of \object{LS~5039} on 1906 in Hipparcos
Celestial Reference System (HCRS, coincident with J2000.0). USNO-A2.0 contains a
global reduction of the data in the Palomar Observatory Sky Survey plates
obtained around 1950. The Twin Astrographic Catalog version 2 (TAC~2.0) provides
a position based on four observations (the plates were compiled between 1977 and
1986). Following \cite{ribo02}, we included an additional uncertainty of 15~mas
in each coordinate, as suggested in \cite{zacharias99}. We also used the Guide
Star Catalog (GSC) version 1.2, which consists of a single-epoch collection of
6.4$^{\circ}\times6.4^{\circ}$ Schmidt plates (the newer versions of the catalog
are compiled, and proper motions are fitted using Tycho positions). The GSC
position is less relevant for the fit because of the relatively large errors
(300~mas) for a position obtained in 1986. The Tycho-2 Catalog was obtained from
the Tycho star mapper observations of the ESA Hipparcos satellite \citep{hog00}.
It provides the mean position at epoch J2000, corrected from proper motion, as
well as the observed position, which is the one used in our analysis. Following
\cite{ribo02}, we have used the conservative error estimate of
$\sigma_{\alpha}\cos\delta=149$~mas and $\sigma_{\delta}=120$~mas.
\cite{skrutskie06} detected \object{LS~5039} at near-infrared bands with the
all-sky Two Micron All--Sky Survey (2MASS), and a very precise position was
obtained. The three observed bands ($J$, $H$, and $K_{\rm s}$) were tied with
the Tycho-2 Reference Catalog, so it is in the ICRS frame. Finally, accurate
astrometry of \object{LS~5039} can be found in the latest versions of the US
Naval Observatory CCD Astrograph Catalog \citep{zacharias10}. However, we can
only use the position from the UCAC1 version \citep{zacharias00}. All these
optical positions and their uncertainties are quoted in
Table~\ref{table:ls_positions}.

\subsubsection{Archival radio astrometry of LS~5039} \label{radio_ls5039}

The binary system \object{LS~5039} is a known radio emitter discovered by
\cite{marti98}. Since then, several radio interferometric observations have been
conducted to study its morphological and spectral properties at different
orbital phases. Here we list the radio projects in which the system has been
observed with sufficient resolution to provide an accurate position of the
source. In particular, it has been observed with the VLA, the VLBA and the EVN
at frequencies between 5 and 15~GHz. Observations at lower frequencies with the
VLA are excluded because they provide positions with uncertainties above
$0\farcs1$ for relatively recent epochs, and do not contribute to the proper
motion fit when compared to the 1--10~mas accuracy of the other radio
observations. 

We briefly summarise the observations included in the fit. \cite{marti98}
conducted a multi-epoch and multi-frequency project of the system with the VLA
at four frequencies between 1 and 15~GHz. They provided a single position
derived from the 15~GHz data obtained with the highest resolution configuration
(A) of the VLA. The reference source was the ICRF source J1911-2006 (1911-201),
and their error estimate is 10~mas. The source was next observed with the VLBA
and the VLA for two epochs in June 2000 at 5~GHz \citep{ribo08}. Only one of the
two epochs provides good astrometry. The source was also observed with the EVN
within project EF018 in March 2007. The project consists of three observations
at 5~GHz, observed every two days. The low elevation of \object{LS~5039} during
the observations affects the phase calibration of the data and therefore the
astrometry. To obtain a reliable position, we removed the data with the lowest
elevations and then averaged the three measurements. The uncertainty of this
position was computed as the standard deviation of the three runs. We conducted
a similar project, observing the source with the VLBA during five consecutive
days in July 2007 at 5~GHz \citep[see details in][]{moldon12vlbi}. A similar
procedure was used in this case, computing the average position of the source
from the five images produced with uniform weighting. Finally, we used a recent
position obtained within the VLBA large project
BC196\footnote{https://science.nrao.edu/observing/largeproposals/blackholesearch/}
(PI: J. Condon), whose solutions are available online. The positions,
uncertainties (see the overview in \citealt{condon11}), and residuals with
respect to the final proper motion fit are shown in
Table~\ref{table:ls_positions}.

\subsection{Uncertainties in the position of LS~5039} \label{uncertainties_ls}

\object{LS~5039} is known to be extended at mas scales, and a bipolar jet-like
structure is present in all VLBI (VLBA and EVN) radio observations of the source
\citep{paredes00, paredes02,ribo08,moldon11_heeps,moldon12vlbi}. Although the
source displays a bright core, we do not know the offset between the peak of the
radio core and the position of the binary system. There are two main reasons.
First, the particles accelerated in the system travel distances much longer than
the system orbit, and the location of the peak of the emission at any moment
depends on the physical conditions (basically the electron distribution, and the
synchrotron cooling time). Second, for each observed frequency, each
interferometer array is sensitive to different ranges of spatial scales,
therefore inter-comparison is not straightforward because we are seeing
different regions of the particle flow. In summary, the unknown position of the
system within the radio outflow is variable (at time scales of 1 or more days),
and depends on the observed frequency and array. We tried to minimise this
effect by measuring positions derived from the uniformly weighted images, which
trace the most compact emission from the core component of the source. 

Preliminary fits of the position and proper motion of \object{LS~5039} using the
formal astrometric errors of the radio positions yielded reduced $\chi^{2}$ of
4.1 and 2.3 in right ascension and declination, respectively, which indicated
that a linear model did not completely describe the source positions. This means
that the formal astrometric errors for the radio observations underestimate our
real knowledge of the position of the binary system within the radio emission. A
shift on the peak of the emission was observed in \object{LS~I~+61~303},
\object{PSR~B1259$-$63}, and \object{HESS~J0632+057} \citep{dhawan06,
moldon11_psr, moldon11_hess}. Based on these observations, and the estimations
in \cite{dubus06} and \cite{bosch-ramon11}, we expect the peak of the emission
to follow an ellipse with semimajor axis of about $\sim$1~mas, although the size
should depend on the frequency, and should be larger at low frequencies. It is
not possible to measure this unknown deviation with the current resolution and
data sampling, especially when observing at different frequencies and
resolutions. Therefore, we increased the position uncertainty of the radio
positions to compensate our ignorance of this time-dependent offset. In
particular, we added in quadrature an {\it ad-hoc} uncertainty of 1.0~mas to all
radio positions before performing the fit. This uncertainty is already included
in Table~\ref{table:ls_positions}.

Another source of uncertainty is the systematic errors on the position of the
phase calibrators, which are the reference of the measured positions.
Fortunately, the phase reference calibrator for all VLBI observations was
\object{J1825$-$1718}. Although the calibrator was correlated at different
positions, we have shifted all measurements to the common reference position
quoted in Sect.~\ref{obs_psr_astrometry}. We followed the same procedure as with
\object{PSR~J1825$-$1446}, and these uncertainties were not included in the
proper motion determination, but were added afterwards to the final fitted
position.

We also included the effect of the annual parallax of \object{LS~5039}, assuming
a distance of 2.9~kpc, which corresponds to a deviation of 0.3~mas. The
equations governing the parallax displacements can be found, for example, in
\cite{brisken02} and \cite{loinard07}. The additional displacements were added
to the data before performing the fit, and are not included in
Table~\ref{table:ls_positions}. 

\subsection{Proper motion of LS~5039} \label{movprop_ls5039}

\begin{figure*}[] 
\begin{center}
\resizebox{1.0\hsize}{!}{\includegraphics[angle=0]{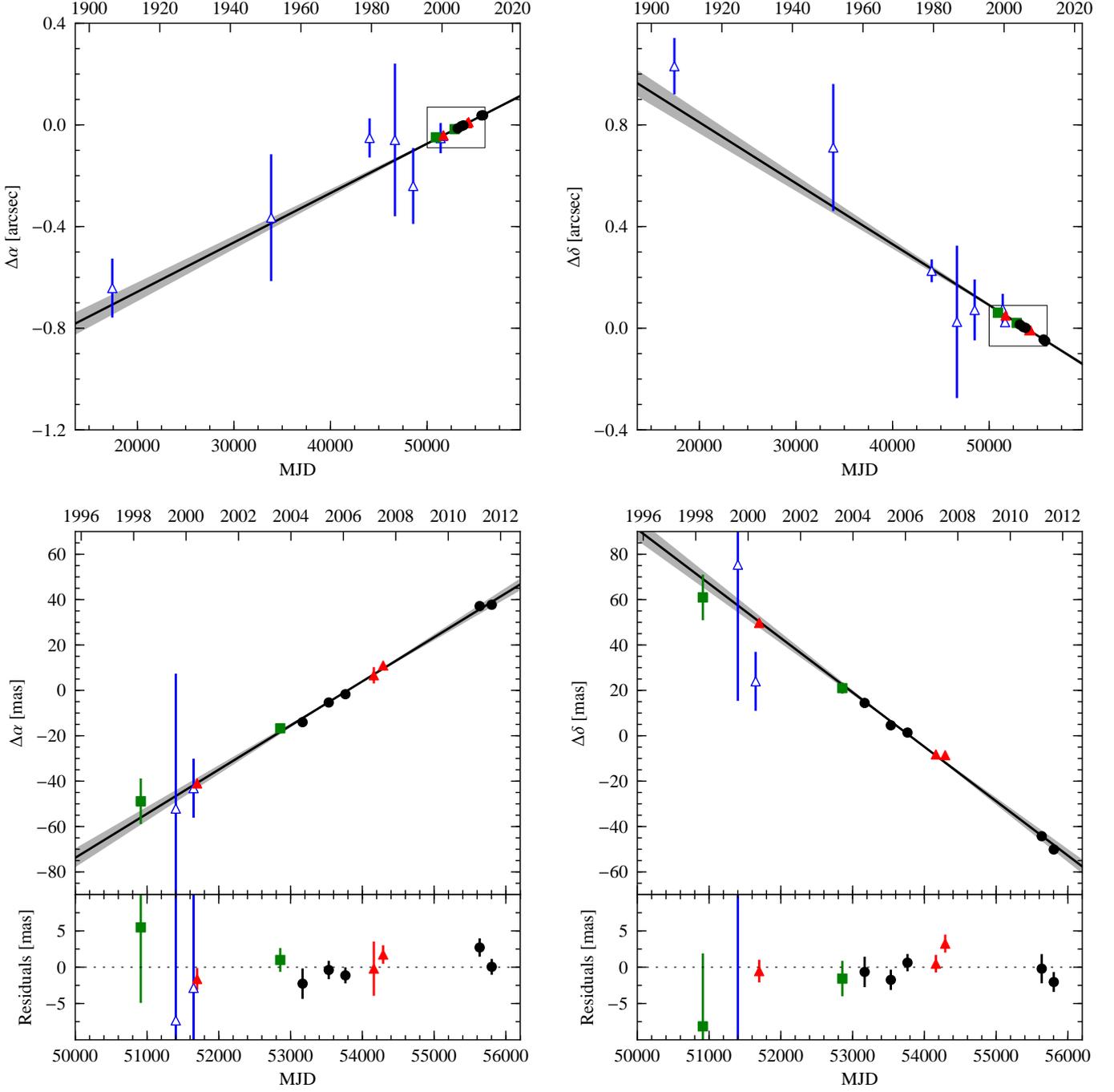}}
\caption{Offsets in right ascension (\emph{left}) and declination (\emph{right})
of \object{LS~5039} with respect to time in MJD and year. The solid black lines
in each panel represent the fitted proper motion, and the grey area the
uncertainty of the fit at 3-$\sigma$ level. The upper panels include all data
points used for the fit, and the lower panels are the zoom of the last 16~years
(marked by the small rectangles), and include the residuals with respect to the
fitted model. The empty blue triangles represent the optical data, the green
squares show the VLA data, the red triangles correspond to the VLBI observations
at 5~GHz, and the black circles are the VLBI data at 8.6~GHz. The source
positions are plotted with uncertainties at 1-$\sigma$ level.
\label{fig:pm_ls}}
\end{center}
\end{figure*}

We fitted the reference position and the proper motion to the data in right
ascension and declination independently for the 16 positions available, using
the Levenberg-Marquardt nonlinear least-squares fit. The reference epoch for the
fit, computed as the weighted mean of the dates of observations, is
MJD~53797.064. The resulting astrometric parameters for \object{LS~5039} are

\begin{align*}
\alpha_\mathrm{J2000}   & =  18^{\rm h}\, 26^{\rm m}\, 15\fs05927\, \left(\pm0.5\pm0.9~\mathrm{mas}\right),     \\
\delta_\mathrm{J2000}   & = -14^{\circ}\, 50^{\prime}\, 54\farcs3010\, \left(\pm0.5\pm1.4~\mathrm{mas}\right),  \\
\mu_\alpha \cos{\delta} & =  7.10 \pm 0.13\ {\rm mas\ yr}^{-1}, \\
\mu_\delta              & = -8.75 \pm 0.16\ {\rm mas\ yr}^{-1},   \\
\end{align*}

\noindent
where the first set of errors in the positions corresponds to statistical
uncertainties, and the second set corresponds to the uncertainty in the phase
reference position. The reduced $\chi^{2}$ of the right ascension and
declination fits were 0.75 and 1.35, respectively. The positions, fit, and
uncertainties are shown in Fig.~\ref{fig:pm_ls}. The solution is dominated by
the radio observations, but the optical observations provide better stability to
the fit. We note that the effect of the parallax correction on the data is
small. If the correction is not included, the reduced $\chi^{2}$ in right
ascension increases by 0.02, and does not change in declination. We also tried
to fit the parallax to the data, with and without the {\it ad hoc} 1~mas
uncertainty, but the parallax is too small to be measured with the current data.

\begin{figure}[h] 
\begin{center}
\resizebox{1.0\hsize}{!}{\includegraphics[angle=0]{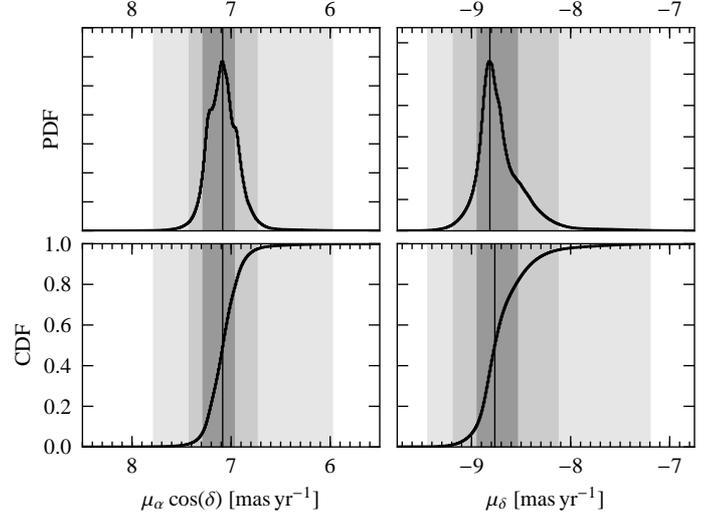}}
\caption{\emph{Top:} distribution of proper motion fit results for LS~5039
estimated with the bootstrap method, which is equivalent to an un-normalised
probability distribution function (PDF) for right ascension and declination
independently. \emph{Bottom}: corresponding cumulative distribution functions
(CDF). For all panels, the grey shaded areas correspond to the minimum region
that includes the 68.2, 95.4 and 99.7\% of the events, from dark grey to light
grey. The vertical line marks the most common value (the mode) for the PDF
distributions, and the value covering the 50\% of the events for the CDF.
\label{fig:ls_distributions}}
\end{center}
\end{figure}

To better understand the systematic fit uncertainties, we analysed the
distribution of fit solutions following a procedure similar to the one discussed
in \cite{chatterjee09}. The bootstrap method used consists of fitting the proper
motion and reference position from different samples of the available positions.
For each iteration we randomly selected 16 positions, with replacement, and
produced a proper motion fit. By repeating the fit for different combinations of
positions, we obtained a probability distribution of proper motions obtained
with all possible combinations of the data. The probability distribution
functions (PDF) and the cumulative distribution functions (CDF) of these PDFs
are plotted in Fig.~\ref{fig:ls_distributions}. The distributions were computed
with a bin width of 0.01~mas~yr$^{-1}$. The number of iterations (different
combinations with replacement) is $2\times10^{7}$, which is more than necessary
to smoothly cover the one-dimensional range of values, but it is required for
the 2D distribution (see below). The shaded grey areas in
Fig.~\ref{fig:ls_distributions} indicate the minimum regions containing the
68.2, 95.4 and 99.7\% of the samples, from darker to lighter, respectively. The
best-fit value of the proper motion can be computed in two ways: as the mode of
the distribution (value that occurs most frequently), which is marked as the
vertical line in the top panels, or as the value that accumulates the 50\% of
the events (the median), which is marked as the vertical line in the bottom
panels of Fig.~\ref{fig:ls_distributions}. These two values are not identical,
but the difference is well below the 1-$\sigma$ uncertainty of the proper motion
estimation. We can clearly see that the distribution is approximately Gaussian
at the central region, but it shows considerable asymmetric wings. The mode and
the confidence intervals are $\mu_{\alpha}\cos{\delta} =
7.09^{+0.19}_{-0.11}$~{mas~yr}$^{-1}$ and $\mu_{\delta} =
-8.82^{+0.27}_{-0.12}$~{mas~yr}$^{-1}$.

\begin{figure}[h!] 
\begin{center}
\resizebox{1.0\hsize}{!}{\includegraphics[angle=0]{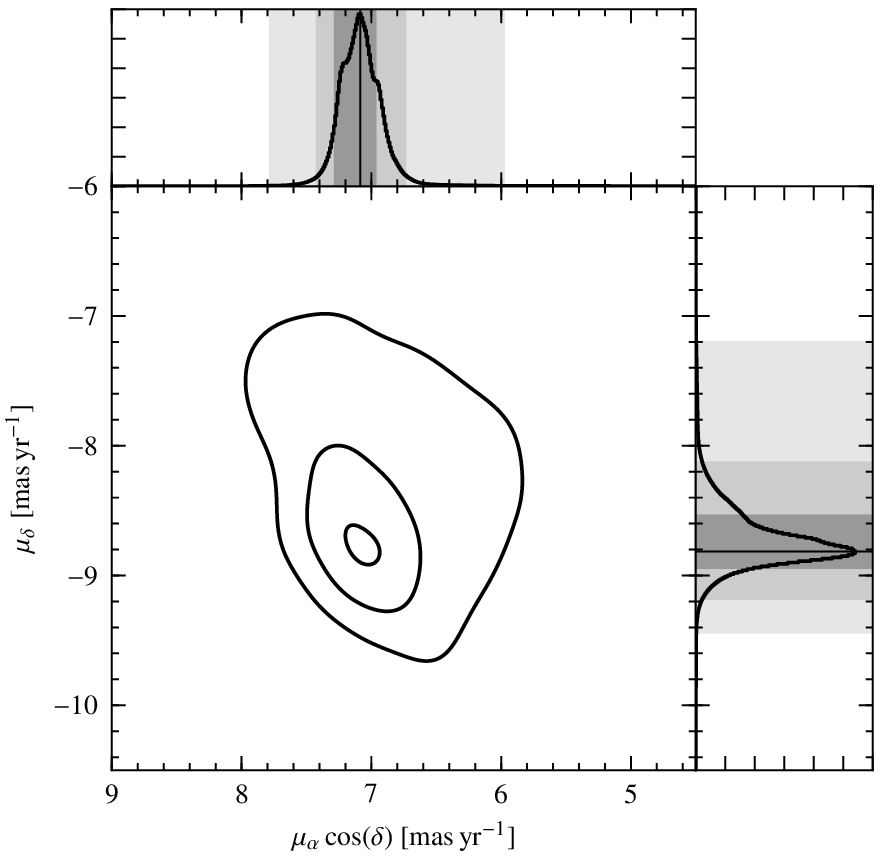}}
\caption{Distribution of proper motion fits for LS~5039 computed with the
bootstrap method. The central contours represent, from inside to outside, the
minimum smoothed regions containing 68.2, 95.4, and 99.7\% of the events. The
two histograms are the projected 1D distributions shown in
Fig.~\ref{fig:ls_distributions}.
\label{fig:ls_contour_distribution}}
\end{center}
\end{figure}

Owing to the asymmetry in the distributions of Fig.~\ref{fig:ls_distributions}
we computed the 2D distribution of proper motions obtained with the bootstrap
method in right ascension and declination simultaneously to better identify any
other correlation or asymmetry between the confidence regions. The simulated fit
results were gridded with a bin size of 0.01~mas~yr$^{-1}$ in both directions.
The minimum regions covering the 68.2, 95.4 and 99.7\% of the events were
obtained, and they were smoothed with a Gaussian filter with a size of 18~bins.
The corresponding confidence regions are shown in
Fig.~\ref{fig:ls_contour_distribution}. We can see that the distribution is
clearly skewed towards lower values of $\mu_{\delta}$ and more moderately to
lower values of $\mu_{\alpha}\cos\delta$.

\section{Galactic space velocity} \label{gal_vel}

We can now study the Galactic space velocity of \object{LS~5039} and
\object{PSR~J1825$-$1446}, following the work in \cite{ribo02}, \cite{dhawan07},
and \cite{miller-jones09}. The space velocity of each source can be obtained by
combining the proper motion and the radial velocity of the source (optical
systemic velocity, $\gamma$, in velocity lightcurves) to form a space velocity
vector $(\mu_{\alpha}\cos\delta, \mu_{\delta}, \gamma)$. This velocity can be
rotated to the cartesian coordinates $U$, $V$, $W$, defined as $U$ positive
towards the Galactic Centre, $V$ positive towards $l = 90^{\circ}$ (sense of the
Galactic rotation), and $W$ positive towards the North Galactic Pole (NGC). The
rotation is described in \cite{johnson87}, although we used the updated
definition of the North Galactic Pole direction from \cite{reid04} (the
direction of the NGP is $\alpha^{\rm J2000.0}_{\rm P}=12^{\rm h} 51^{\rm m}
26\fs282$ and $\delta^{\rm J2000.0}_{\rm P}=+27\degr 07\arcmin 42\farcs01$, with
the zero longitude at $\theta^{\rm J2000}_{\rm 0}=122\fdg932$). We computed the
heliocentric Galactic space velocity by subtracting the peculiar velocity of the
Sun with respect to the local standard of rest (LSR). We used the recent
determination from \cite{schonrich10} of $U_{\odot}=11.1^{+0.69}_{-0.75}$,
$V_{\odot}=12.24^{+0.47}_{-0.47}$, $W_{\odot}=7.25^{+0.37}_{-0.36}$~km~s$^{-1}$.
We note that $V_{\odot}$ is significantly larger than in previous estimations.
The obtained heliocentric Galactic velocities $(U, V, W)_{\rm LSR}$ for
\object{LS~5039} and \object{PSR~J1825$-$1446} are shown in
Table.~\ref{table:uvw}. 

\begin{table} 
\begin{center}
\caption{Galactic coordinates and Galactic space velocities of LS~5039, and PSR~J1825$-$1446. In the Galactic velocities, the first set of errors corresponds to the proper motion uncertainty, while the second set (in parentheses) corresponds to the distance uncertainty.}
\label{table:uvw} 
\begin{tabular}{l r r}
\midrule
\midrule
Parameter                                   &  LS~5039                       & PSR~J1825$-$1446     \\
\midrule
Galactic Longitude $l$                      &  16\fdg882                     &  16\fdg805           \\         
Galactic Latitude $b$                       &  -1\fdg289                     &  -1\fdg001           \\         
Distance $d$ (kpc)                          & $2.9\pm0.8$                    & $5.0\pm0.6$          \\        
\midrule
$\mu_{\alpha}\cos{\delta}$ (mas~yr$^{-1}$)  & $7.09^{+0.19}_{-0.11}$         & $10.0\pm0.3$         \\
$\mu_{\delta}$ (mas~yr$^{-1}$)              & $-8.82^{+0.27}_{-0.12}$        & $-29.0\pm0.3$        \\
Sys. velocity ($\gamma$) (km~s$^{-1}$)      & $17.2\pm0.5$                   & --                   \\    
$\mu_{\rm l}$ (mas~yr$^{-1}$)               & $-4.50^{+0.25}_{-0.12}$         & $-21.0\pm0.3$       \\
$\mu_{\rm b}$ (mas~yr$^{-1}$)               & $-10.38^{+0.21}_{-0.11}$       & $-22.4\pm0.3$        \\
\midrule
$U_{\rm LSR}$ (km~s$^{-1}$)                 & $42.5^{+1.0}_{-1.3}~(\pm4)$    & $146\pm2~(\pm16)$    \\       
$V_{\rm LSR}$ (km~s$^{-1}$)                 & $-43^{+3}_{-1.6}~(\pm17)$      & $-466\pm7~(\pm60)$  \\       
$W_{\rm LSR}$ (km~s$^{-1}$)                 & $-136^{+2}_{-2}~(\pm40)$       & $-523\pm7~(\pm60)$  \\      
\midrule
$v_{\rm rad}$ (km~s$^{-1}$)                 & $-11.2^{+1.1}_{-1.4}~(\pm5)$   & $227\pm3~(\pm40)$   \\        
$v_{\rm cir}$ (km~s$^{-1}$)                 & $197^{+3}_{-1.6}~(\pm17)$      & $-150\pm6~(\pm60)$  \\       
$v_{\rm z}$ (km~s$^{-1}$)                   & $-136^{+2}_{-2}~(\pm40)$       & $-523\pm7~(\pm60)$  \\       
$v_{\rm pec}$ (km~s$^{-1}$)                 & $142^{+2}_{-2}~(\pm40)$        & $690\pm7~(\pm60)$  \\       
\midrule
\end{tabular}
\end{center}
\end{table}

We note that the distance to the source is the highest source of error when
converting from angular to linear velocities. We used the most recent
determination of the distance to \object{LS~5039}, $2.9\pm0.8$~kpc
\citep{casares12_ls}. The distance to \object{PSR~J1825$-$1446} is
$5.0\pm0.6$~kpc. The first set of errors in the Galactic velocities in
Table~\ref{table:uvw} corresponds to the proper motion uncertainty, and the
second set of errors corresponds to the contribution of the distance
uncertainty.

We can now compute the Galactic space velocity as measured from the Galactic
Centre (GC). We need to transform the velocity vector to the reference frame of
an observer situated at the position of the source that participates in the
Galactic rotation. This reference frame is the regional standard of rest (RSR).
The transformation from the LSR to the RSR takes into account the distance from
the Sun to the GC, $8.0\pm0.5$~kpc \citep{reid93}, and the circular rotation
about the GC, 236~km~s$^{-1}$ \citep{reid04}, which we consider constant. From
the Galactic velocity $(U, V, W)_{\rm RSR}$, we directly obtain the
Galactocentric velocity of a source, with the radial velocity $v_{\rm rad}$ away
from the GC, $v_{\rm cir}$, in the direction of the Galactic rotation at each
point, and $v_{\rm z}$, perpendicular to the Galactic Plane, and positive
towards the NGP. The expected velocity of a system following the Galactic
rotation is zero for $v_{\rm rad}$ and $v_{\rm z}$, whereas $v_{\rm cir}$ is
expected to be 236~km~s$^{-1}$. The deviation from the expected circular
rotation is the peculiar velocity $v_{\rm pec}= [v^{2}_{\rm rad} + (v_{\rm
cir}-236{\rm ~km~s}^{-1}) ^{2} + v^{2}_{\rm z} ]^{1/2}$. The Galactic velocities
from the GC for \object{LS~5039} and \object{PSR~J1825$-$1446} are shown in
Table.~\ref{table:uvw}.

\begin{figure}[] 
\begin{center}
\resizebox{1.0\hsize}{!}{\includegraphics[angle=0]{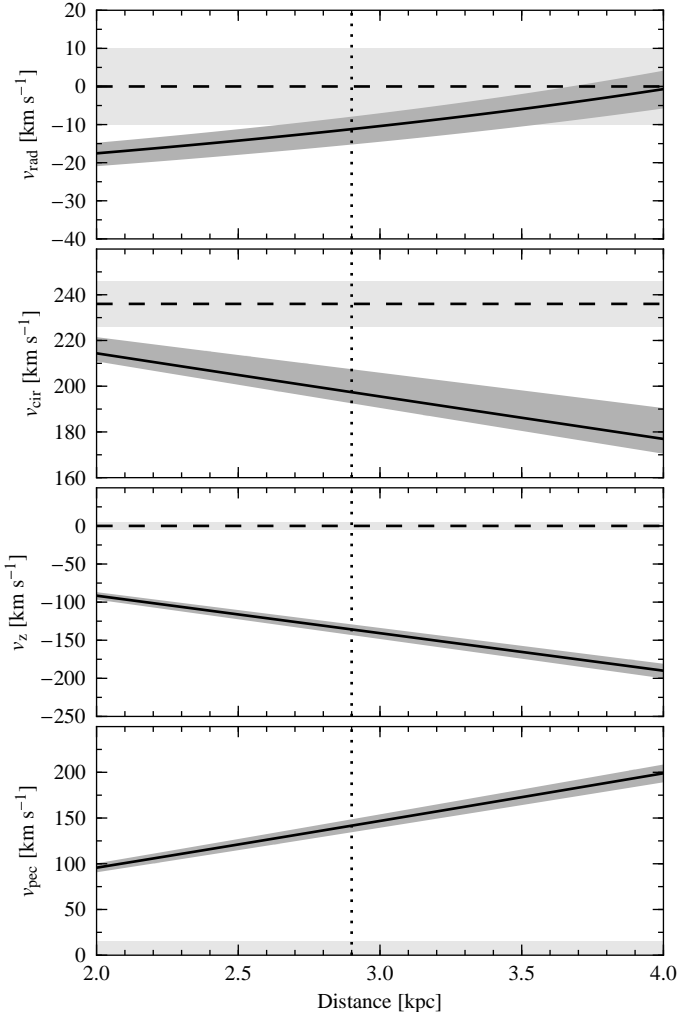}}
\caption{Components of the Galactic and peculiar velocity of LS~5039 (solid
black lines) and their uncertainties at 3-$\sigma$ level (dark grey areas). The
horizontal dashed lines and light grey areas mark the expected Galactic
velocity, and the velocity dispersion expected for young stars, respectively.
The vertical dotted lines indicate the measured distance to the source.
\label{fig:gal_vel_ls}}
\end{center}
\end{figure}

\begin{figure}[] 
\begin{center}
\resizebox{1.0\hsize}{!}{\includegraphics[angle=0]{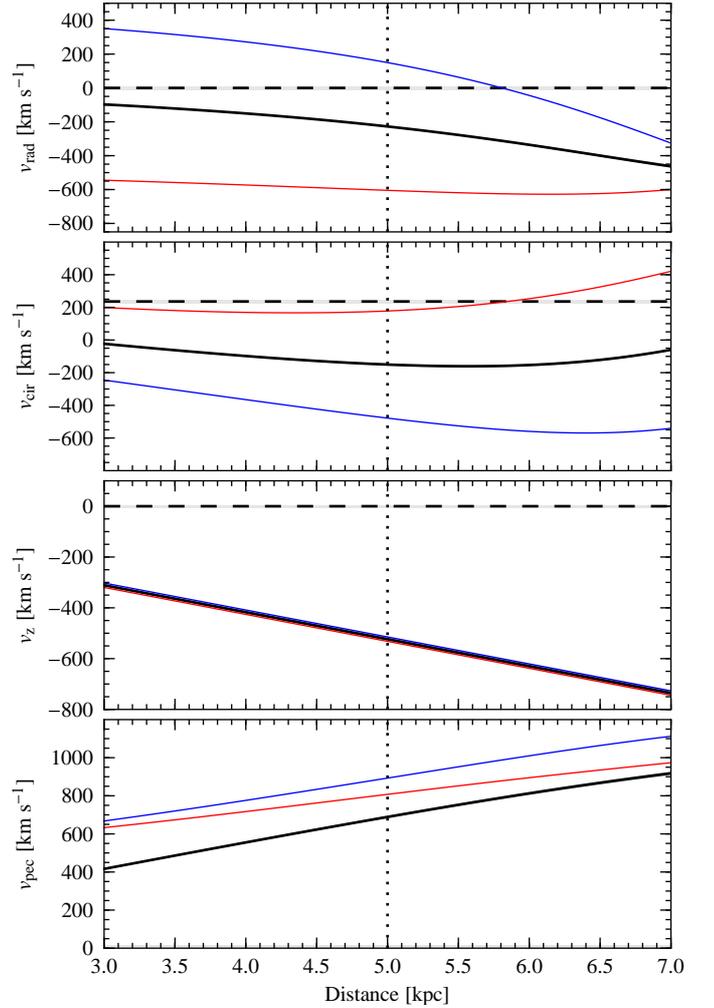}}
\caption{Same as Fig.~\ref{fig:gal_vel_ls} for PSR~J1825$-$1446. The solid black
line in each panel indicates the Galactic velocity assuming a radial velocity of
0~km~s$^{-1}$. The velocity uncertainties at 3-$\sigma$ level are smaller than
the black lines. The red thin line corresponds to an hypothetical radial
velocity from the Sun of $+500$~km~s$^{-1}$ (away from the Sun) and the blue
thin line to $-500$~km~s$^{-1}$ (towards the Sun). 
\label{fig:gal_vel_psr}}
\end{center}
\end{figure}

The large error in the distance of the sources limits the determination of their
space velocity. In this case it is convenient to plot the Galactic velocities as
a function of the distance from the Sun. In Figs.~\ref{fig:gal_vel_ls} and
\ref{fig:gal_vel_psr} we show the computed velocities (black lines) and their
uncertainties (dark grey areas) for a range of distances. The horizontal dashed
lines mark the expected velocity when following the Galactic rotation. The grey
areas indicate the cosmic velocity dispersion of the Galactic young disc stars,
which depends on the age of the stellar system. \object{LS~5039} is an O-type
star and its companion could have been more massive. The progenitor of
\object{PSR~J1825$-$1446} is unknown, but collapse to form a neutron star
usually occurs for stars with total mass between 8 and 15~M$_{\odot}$
\citep{lyne05}. Therefore, we assumed that the progenitor was a young star
formed in the Galactic thin disc. The measured velocity dispersion of young OB
stars is biased towards local values because the stars with known 3D velocities
and radial distances are predominantly located in the solar neighbourhood.
However, the velocity dispersion of young stars should follow the velocity
dispersion of the gas in the Galactic disc. This dispersion is of about
4--5~km~s$^{-1}$ in the plane $(\sigma_{U}, \sigma_{V})$, and can be as low as
1--2~km~s$^{-1}$ in the direction perpendicular to the plane $(\sigma_{W})$,
although spiral shocks can induce higher in-plane velocities of about
$\sim7-10$~km~s$^{-1}$ \citep{kim06}. This agrees with values measured in
\cite{torra00} for stars of 0--30~Myr of $(\sigma_{U}, \sigma_{V}, \sigma_{W}) =
(7.9, 7.2, 4.3)$~km~s$^{-1}$. We assume that the expected velocity dispersion of
young stars is $(\sigma_{U}, \sigma_{V}, \sigma_{W}) = (10, 10, 5)$~km~s$^{-1}$.

We can see in Fig.~\ref{fig:gal_vel_ls} that \object{LS~5039} has a high
peculiar velocity, mainly perpendicular to the Galactic plane, and it is not
compatible with the expected velocity dispersion for any of the considered
distances. This reaffirms the runaway nature of the system, which was originally
discussed in \cite{ribo02}. On the other hand, \object{PSR~J1825$-$1446} shows a
much higher space velocity. This is an isolated pulsar, and therefore it is not
possible to measure any radial velocity from the Sun. In
Fig.~\ref{fig:gal_vel_psr} we show the Galactic velocities assuming a radial
velocity of 0, +500, and $-500$~km~s$^{-1}$ (black, red, and blue lines,
respectively). This covers a generous range of possible radial velocities. We
note that this contribution is less notorious in the direction perpendicular to
the Galactic Plane because both the pulsar and the Sun are at a similar distance
from the Galactic plane. For \object{PSR~J1825$-$1446} we plot a wide range of
distances, considering that the distance estimate could be overestimated by the
DM determination. The pulsar space velocity is not compatible with the expected
rotation of the Galaxy. Remarkably, the pulsar has a velocity perpendicular to
the Galactic Plane of $\sim550$~km~s$^{-1}$ and away from it. Therefore,
\object{PSR~J1825$-$1446} is clearly a runaway pulsar that has been ejected from
the Galatic disc at a very high speed. The peculiar velocity for the nominal
distance of 5.0~kpc is $690\pm7$~km~s$^{-1}$. This transverse velocity is high
even for a runaway isolated pulsar, because it lays in the tail of the
transverse velocity distribution of young pulsars shown in \cite{hobbs05}. The
peculiar velocity would be lower if the distance is overestimated, and would be
$500\pm5$~km~s$^{-1}$ for a distance of 3.6~kpc.

\section{The origin of LS~5039 and PSR~J1825$-$1446} \label{origin}

\object{LS~5039} is a young O6.5 main sequence star, which has a total lifetime
of a few million years. This sets an upper limit to the age of the binary system
\object{LS~5039}/compact object, which allows us to constrain the possible
locations where the supernova took place, and therefore where the system
originated. We have explored possible origins of \object{LS~5039} in the last
1--10~Myr, even though for ages older than 1--2~Myr the source birthplace lies
well outside the Galactic plane, which is unlikely. Therefore, we only consider
the projected past trajectory of \object{LS~5039} during the last
$\sim10^{6}$~yr, and discuss possible associations that cross this trajectory.
We know that the characteristic age of \object{PSR~J1825$-$1446} is
1.95$\times$10$^5$~yr. This value is computed as the ratio between the period
and the period derivative of the pulsar, and it is valid under certain
assumptions: the initial spin period of the pulsar was much shorter than that
observed today ($P_{0}\ll P$), there is no magnetic field decay ($P\dot{P}$ is
constant), and the energy loss corresponds to a spinning dipolar magnet in
perfect vacuum. The slowdown model of a pulsar is generally parametrised by the
braking index, which is usually considered close to 3, although it can take a
range of values (see \citealt{faucher06}). The uncertainty on this parameter
prevents the characteristic age from being a faithful estimate. It has been
shown that the characteristic age of a pulsar can be significantly lower than
the true age of the pulsar (e.g. \citealt{gaensler00}) or even higher than the
real age of the pulsar (e.g. \citealt{kramer03}). Therefore, a wide range of
ages, and consequently possible angular distances from the current location of
the pulsar, should be explored.

\begin{figure*}[] 
\begin{center}
\resizebox{1.0\hsize}{!}{\includegraphics[angle=0]{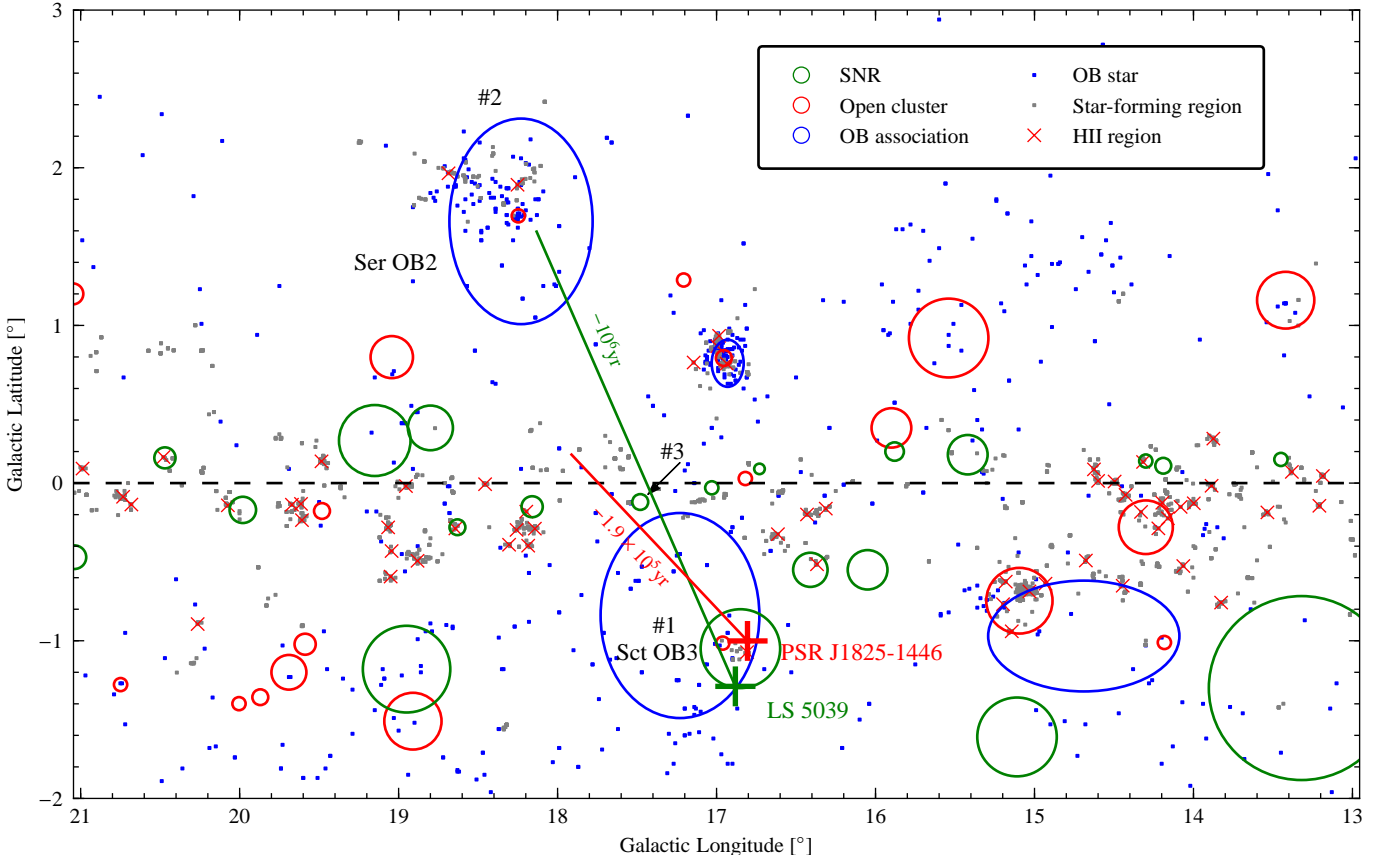}}
\caption{Area around LS~5039 and PSR~J1825$-$1446 in Galactic coordinates. The
lines mark the past trajectory in the last $10^{6}$~yr for LS~5039, and
$1.95\times10^{5}$~yr for PSR~J1825$-$1446. The green circles are supernova
remnants from the catalogue compiled in \cite{green09}. The red circles are open
clusters from the WEBDA database, and the blue ellipses mark the position of the
known OB associations \citep{melnik95}. The OB stars \citep{reed03} and the
star-forming regions \citep{avedisova02} are marked with blue and grey dots,
respectively, and provide a general view of the young system distribution in
this region of the Galaxy. The red crosses mark the \ion{H}{ii} regions
\citep{lockman89}. The objects in regions \#1, \#2, and \#3 are described in the
text.
\label{fig:origin}}
\end{center}
\end{figure*}

In Fig.~\ref{fig:origin} we show the region around the current position of
\object{LS~5039} and \object{PSR~J1825$-$1446} (green and red crosses,
respectively) in Galactic coordinates. We also plot different objects related
with young stars. The ellipses in green, red, and blue mark the positions of the
supernova remnants, the open clusters and the OB associations, respectively. To
have a better context of the distribution of young systems, we include the O and
B stars, the star-forming regions, and the \ion{h}{ii} regions close in this
part of the Galactic plane. The past trajectories in the last $10^{6}$~yr for
\object{LS~5039}, and $1.95\times10^{5}$~yr for \object{PSR~J1825$-$1446} are
indicated by the green and red continuous lines, respectively. The plotted
trajectories are linear, and therefore not affected by the Galactic potential.
We checked the validity of this assumption using the software package
\emph{galpy}\footnote{https://github.com/jobovy/galpy} to compute the
trajectories under a standard Milky Way potential (combination of a
Miyamoto-Nagai disc, a Navarro-Frenk-White halo, and a Hernquist bulge
potentials, see the documentation for details). For all aspects discussed in
this paper the slight difference between the integrated orbit in the Galactic
potential and the free (linear) trajectory are irrelevant. Therefore, we do not
use the orbit integration to avoid using additional parameters to describe the
Galaxy. We used the software to check that the space velocity of
\object{PSR~J1825$-$1446} is above the escape velocity, and therefore it will be
ejected from the Galaxy.

As can be seen in Fig.~\ref{fig:origin}, the past trajectory of \object{LS~5039}
and \object{PSR~J1825$-$1446} crosses three interesting regions, marked with
numbers from \#1 to \#3. The first region (\#1) is composed of the objects at
short angular distances from the systems. In the WEBDA
database\footnote{http://www.univie.ac.at/webda/} we can see that
\object{LS~5039} and \object{PSR~J1825$-$1446} are within the open cluster
\object{Dolidze~28} (C~1822$-$146), whose central star is \object{WR~115}.
\cite{manchanda96} determined the distance to the cluster to be 2.2~kpc (note
that the cluster is wrongly named Do~78 instead of Do~28 in Table~1 of
\citealt{manchanda96}). We used the 2MASS photometric magnitude of
\object{WR~115} (Wolf-Rayet of type WN6) from Table~A1 in \cite{crowther06},
$M_{\rm Ks}=-5.10$, and colours for a nominal subtype WN6 star $(J-K)_{\rm 0}=
0.17$ and $(H-K)_{\rm 0}= 0.15$. The absorption estimate of the source is
$A_{\rm K} = 0.59$, which yields a distance of 2.0~kpc, compatible with the
distance quoted above. In the same field we find the OB association
\object{Sct~OB3} (see \citealt{melnik95}), with ten members in a region of
25.8$\times$33.6~pc$^{2}$, and an estimated distance of 1.5~kpc. This distance
is compatible with the estimate in \cite{tetzlaff10}, obtained from a parallax
measurement, of 1.4~kpc. They also provide an age of \object{Sct~OB3} of
4--5~Myr. The closest \ion{H}{ii} ionised region is \object{RCW~164}, already
discussed in Sect.~\ref{ls_snr}.

\begin{figure}[] 
\begin{center}
\resizebox{1.0\hsize}{!}{\includegraphics[angle=0]{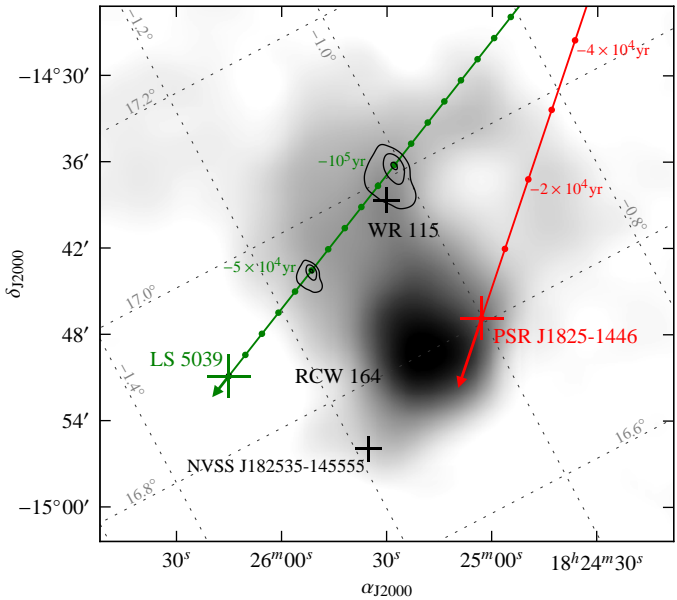}}
\caption{Wide-field map around SNR~G016.8$-$01.1. The grey scale corresponds to
the Parkes-MIT-NRAO survey at 5~GHz \citep{tasker94}. The green and red lines
mark the past trajectory of LS~5039 and PSR~J1825$-$1446, respectively. The
arrows mark the proper motion sense. Circles are plotted in the trajectories
every $10^{4}$~yr. The contours represent the 68.2, 95.4 and 99.7\% confidence
levels on the position that LS~5039 had $10^{5}$ and $5\times10^{4}$~years ago
(see uncertainties in the proper motion shown in
Fig.~\ref{fig:ls_contour_distribution}). The grey dotted lines show the Galactic
coordinates.
\label{fig:wide-field}}
\end{center}
\end{figure}

In Fig.~\ref{fig:wide-field} we show the most relevant objects of region \#1
together with the past trajectories of \object{LS~5039} and
\object{PSR~J1825$-$1446}. We can see that although \object{PSR~J1825$-$1446} is
currently very close to \object{RCW~164}, it comes from the opposite direction
at high speed. The trajectory is not compatible with the central region of
SNR~G016.8$-$01.1 either. Although it is not possible to determine the kinematic
centre of the SNR precisely, the projected past trajectory of \object{LS~5039}
seems to be far away from the central region of the remnant (see
Fig.~\ref{fig:intro_wide-field}), although an association cannot be discarded.
We note that the 1-$\sigma$ accuracy in the position of \object{LS~5039}
$10^{5}$~yr ago is approximately 18~arcsec, although the uncertainties do not
follow a Gaussian distribution, as shown by the contours in
Fig.~\ref{fig:wide-field}.

The past trajectory of \object{LS~5039} shown in Fig.~\ref{fig:wide-field}
suggests an interesting possibility, namely that the binary system
\object{LS~5039} and \object{WR~115} come from the same point on the sky, and
that they might have been a triple system, from which \object{WR~115} was
ejected in the opposite direction after the supernova explosion. This would
tightly restrict the age of the triple system. However, we have computed a
proper motion of \object{WR~115} using positions from optical catalogues, in
particular from AC2000.2, GSC~1.2, Hipparcos, UCAC3, and 2MASS, see
Sect.~\ref{optical_ls5039}. The proper motion obtained is $4.5\pm0.6$ and
$-5\pm4$~mas~yr$^{-1}$ in right ascension and declination, respectively. The
proper motion in right ascension discards the possibility of a triple system
with \object{WR~115} because the linear momentum conservation would require that
the system pre-supernova already had a runaway velocity much higher than
expected from Galactic diffusion.

Region \#2 is found at an angular distance of $\sim3^{\circ}$ from both sources,
and harbours an active site of star formation and young stars. The OB
association shown in region \#2 is \object{Ser~OB2} \citep{melnik95}, with a
size of 22.8 and 33.0~pc in Galactic longitude and latitude, respectively, an
average radial velocity relative to the Sun of $-2.4\pm3.9$~km~s$^{-1}$, and an
average distance of 1.45~kpc. \cite{forbes00} provides a distance to the
association of $1.9\pm0.3$~kpc, and an average age of $5\pm1$~Myr.
\cite{tetzlaff10} provide a parallax measurement based on Hipparcos stars of
0.63~mas, which correspond to a distance from the Sun of 1.6~kpc and sets an age
of the association of 4--5~Myr. On the other hand, the open cluster in the
centre of region \#2 is NGC~6604, which can be found in the WEBDA catalogue,
with a distance of 1.7~kpc, and an age of $\sim6$~Myr. \cite{loktin03} measured
an average proper motion of this cluster, formed by $\sim97$ stars, and obtained
$-0.66\pm0.19$ and $-0.88\pm0.17$~mas~yr$^{-1}$ in right ascension and
declination, respectively, as well as the distance quoted in the WEBDA database.
Recently, \cite{tetzlaff10} set a parallax of 0.59~mas, corresponding to a
distance from the Sun of 1.7~kpc, and an age of 4--5~Myr. There is also an
interesting \ion{H}{ii} region that might be associated with this complex,
\object{SH~2-54}, from the catalogue of \ion{H}{ii} regions in
\cite{sharpless59}. Its radial velocity was measured by \cite{blitz82} to be
$27.6\pm0.5$~km~s$^{-1}$, much higher than the average velocity of
\object{Ser~OB2} quoted above, and the distance from the Sun is $2.0\pm0.2$~kpc,
which is compatible within errors with the structures discussed above. In
summary, region \#2, which harbours many young stars and star-forming regions,
is undergoing intense formation of new stars, it is at a distance of
1.5--2.0~kpc from the Sun, and has an age of 4--5~Myr. The age of the complex is
enough to allow \object{LS~5039} to have travelled from that region of the sky
to its current position. Considering the large uncertainties in the distance to
the system, this is a plausible birth region for \object{LS~5039} provided that
it is located between 1.5 and 2.0~kpc from the Sun. For this distance range,
Fig.~\ref{fig:gal_vel_ls} shows that \object{LS~5039} would have a lower
peculiar Galactic velocity of about 70--95~km~s$^{-1}$. In this case,
\object{LS~5039} would have an age between 1.0 and 1.2~Myr.

Finally, region \#3 in Fig.~\ref{fig:origin} marks the position of
\object{SNR~G017.4$-$00.1}, which has a diameter of 6$^{\prime}$, a flux density
of 0.4~Jy, and a spectral index of +0.7 \citep{brogan06, green09}. This partial
shell remnant is of class II, which corresponds to a gravitational collapse
supernova explosion. The minimum projected distance between the past trajectory
of \object{LS~5039} and this SNR took place $4\times10^{5}$~yr ago, and for
\object{PSR~J1825$-$1446} it took place $0.13\times10^{5}$~yr ago. For these
high ages we do not expect to still detect the SNR, consequently it is unlikely
that \object{PSR~J1825$-$1446} and \object{SNR~G017.4$-$00.1} are related,
whereas it is not possible that \object{LS~5039} and \object{SNR~G017.4$-$00.1}
are related. On the other hand, \cite{bochow11} used the $\Sigma-D$ relationship
to estimate the distance to \object{SNR~G017.4$-$00.1}, and obtained 12~kpc. The
distance to the SNR seems considerably larger than the distance to the sources,
although we note that this estimate can be very uncertain.

Besides the specific regions on the sky in the past trajectory of both systems,
we can constrain the lifetime of the systems considering that the systems had
contained young massive stars, which are generally formed in a specific region
of the Galactic plane, the thin disc. The stellar space density decays
exponentially as a function of the distance above or below the Galactic plane,
and each stellar population has a particular scale height $h$. For young OB
stars $h_{\rm OB} = 45$~pc \citep{reed00}, while \cite{maiz01} found that the
O--B5 stellar population is distributed with a scale height of
$34.2\pm0.8\pm2.5$~pc. No good determination for the O-type population alone has
been found, basically due to the lack of good statistics with this population,
and therefore we used the more conservative value of 45~pc both for
\object{LS~5039} and \object{PSR~J1825$-$1446}. From the Galactic latitudes
quoted in Table~\ref{table:uvw}, the current height with respect to the Galactic
plane is $-65$ and $-96$~pc for \object{LS~5039} and \object{PSR~J1825$-$1446},
respectively. With the proper motion of the systems, we compute for different
distances from the Sun the time when they were crossing the Galactic plane at a
height within $\pm45$~pc. The expected ages are quoted in
Table~\ref{table:crossing_time}.

\begin{table} 
\begin{center}
\caption{Crossing times of LS~5039 and PSR~J1825$-$1446 through the Galactic
thin disc, with a scale height for OB stars of $\pm45$~pc, for different
distances from the Sun.}
\label{table:crossing_time} 
\begin{tabular}{cc c cc}
\midrule
\midrule

\multicolumn{2}{c}{LS~5039}  & &  \multicolumn{2}{c}{PSR~J1825$-$1446}  \\
\cmidrule{1-2}     \cmidrule{4-5}  
Distance [kpc] & Age [kyr]  & & Distance [kpc] & Age [kyr]  \\
2.0      &   0--890         & & 3.0      &  20--300   \\
2.9      & 140--750         & & 5.0      &  80--245   \\
4.0      & 220--670         & & 7.0      & 100--220   \\
\midrule
\end{tabular}
\end{center}
\end{table}

\section{Discussion and conclusions} \label{conclusions}

We have obtained a set of accurate positions of the isolated pulsar
\object{PSR~J1825$-$1446} by means of VLBI observations. We used the pulsar
gating technique to enhance the detected pulsar flux density to improve the
astrometry. We fitted a linear proper motion to new and archival astrometry and
obtained a proper motion of $\mu_{\alpha}\cos{\delta} =
10.0\pm0.3$~{mas~yr}$^{-1}$, and $\mu_{\delta} = -29.0\pm0.3$~{mas~yr}$^{-1}$.
For a distance from the Sun of 5.0~kpc, the proper motion of
\object{PSR~J1825$-$1446} corresponds to a projected transverse velocity of
$690\pm7$~km~$^{-1}$, which makes it a high-velocity pulsar ejected from the
Galactic plane.

We note that the high space velocity of \object{PSR~J1825$-$1446} suggests that
it is possible that a bow-shock has formed in front of the pulsar. For example,
the pulsar wind nebula G359.23−0.82 is the bow shock produced by the young
pulsar J1747$-$2958, which has a spin-down luminosity of
$2.5\times10^{36}$~erg~s$^{-1}$, and a transverse velocity of 300~km~s$^{-1}$ in
a medium with hydrogen number density of 1.0~cm$^{-3} $\citep{hales09}. However,
the spin-down luminosity of \object{PSR~J1825$-$1446} is only
$4.1\times10^{34}$~erg~s$^{-1}$, and therefore a such structure would only be
visible if the medium surrounding \object{PSR~J1825$-$1446} were much denser
than in the case of J1747$-$2958.

The space velocity of the source has allowed us to investigate possible origins
of \object{PSR~J1825$-$1446}. The characteristic age of
\object{PSR~J1825$-$1446} is approximately 0.2~Myr. The velocity and age of
\object{PSR~J1825$-$1446} make it incompatible with \object{SNR~G016.8$-$01.1}.
There are no clear OB associations or SNRs crossing the past trajectory of the
pulsar. Our estimate of the age of the pulsar, assuming that its progenitor died
in the Galactic disc, is 80--245~kyr, compatible with its characteristic age. We
have seen that the space velocity of this pulsar is high enough to escape from
the Galactic potential.

We compiled all radio interferometric observations of \object{LS~5039} with
accurate astrometry, and reduced all data in a consistent way. Combining the
radio positions with archival optical astrometry from global catalogues we
obtained a proper motion of the source of $\mu_{\alpha}\cos{\delta} =
7.09^{+0.19}_{-0.11}$~{mas~yr}$^{-1}$, and $\mu_{\delta} =
-8.82^{+0.27}_{-0.12}$~{mas~yr}$^{-1}$. We have reduced the uncertainty on the
proper motion by approximately one order of magnitude with respect to the one
obtained in \cite{ribo02}, which is compatible at 2-$\sigma$ level. The data
were collected from very heterogeneous observations (completely different
instruments, conditions, frequencies, and reference calibrators), and the fit is
affected by systematic errors. We investigated the effects of these unknown
uncertainties using a bootstrap method, which provides realistic uncertainties
of our measurement. We confirmed the runaway nature of the system, which has a
peculiar space velocity of $141^{+4}_{-3}$~km~s$^{-1}$, with a main component of
$-136^{+3}_{-2}$~km~s$^{-1}$ perpendicular to the Galactic plane, for a distance
to the source of 2.9~kpc.

We we unable to identify a secure origin for \object{LS~5039}. We restricted the
proper motion of the system, and the past trajectory is not far from the
putative centre of \object{SNR~G016.8$-$01.1}, which is very uncertain. However,
this association would imply an age of 10$^{5}$~yr for the SNR, which is above
the expected lifetime for such a bright SNR \citep{frail94}. The age of the SNR
depends on the ambient density and its distance, which are unknown parameters
\citep[see][]{ribo02}. However, even if \object{SNR~G016.8$-$01.1} is not
related to \object{LS~5039}, the entire region shown in
Fig.~\ref{fig:wide-field} is an active region of stellar formation, with the
open cluster \object{Dolidze~28}, \object{WR~115}, and the \ion{H}{ii} region
\object{RCW~164}. It is possible that the compact object was formed within this
complex. In this case, \object{LS~5039} should be at a distance of around 2~kpc,
and the age of the binary would be below 0.1--0.2~Myr. On the other hand, the
region of \object{Ser~OB2} lies within the past trajectory of \object{LS~5039},
at a distance from the Sun of 1.5--2.0~kpc, which is still compatible at
2-$\sigma$ level with the distance to \object{LS~5039}, 2.9$\pm$0.8~kpc. In this
case the age of the system would be 1.0--1.2~Myr, and the peculiar Galactic
velocity would be 70--95~km~s$^{-1}$. A third possibility is that the system
\object{LS~5039} was formed in the Galactic plane, but it was not associated
with any known young stellar association. In this case the age of the system
would be between 0.1 and 0.8~Myr, assuming a distance of 2.9~kpc.

The age of \object{LS~5039} has consequences on the nature of the compact
object. When the compact object was formed, the system received a powerful kick,
and therefore the orbit could not be pseudo-synchronised after the explosion.
Pseudo-synchronous rotation is reached through tidal forces in an eccentric
binary, when the rotation of the massive star and the motion of the compact
object at periastron are synchronised \citep{hall86}. \cite{casares05}
determined that the compact object should be a black hole (low system
inclination and high compact object mass) if the orbit of the system is
pseudo-synchronised. They estimated the time for this system to reach orbital
pseudo-synchronism to be $\sim1$~Myr, although it is highly uncertain. If the
system was formed in \object{Sct~OB3} region, its age would be below 0.2~Myr,
and the hypothesis in \cite{casares05} would be unlikely. Only if the source was
formed close to \object{Ser~OB2} region, 1--2~Myr ago, would it be possible to
reach orbital pseudo-synchronism. Based on the proper motion of the source, it
is not expected that the system age is greater than 1--2~Myr if the system was
born in the Galactic disc. 

We have seen that any determination of the age and origin of \object{LS~5039} is
limited by the uncertainty in its distance to the Sun. This distance cannot be
obtained by means of parallax VLBI measurements with the current resolution, and
the effects of the source structure and variability. An accurate distance will
be obtained by the Gaia satellite \citep[see][and references therein]{bruijne12}
through measurements of the parallax of the massive companion. The satellite
will also determine the distance to the OB associations with high accuracy, and
therefore any identification or association should be much more straightforward,
and will help to clarify the origin of the system. On the other hand, the
distance to \object{PSR~J1825$-$1446} could be obtained by means of a
challenging long-term VLBI project. This would provide a model-independent
distance that can be compared to the estimated one from the dispersion measure.
Finally, there is no obvious approach to accurately determine the distance to
\object{SNR~G016.8$-$01.1}, whose radio structure is severely contaminated by
the \ion{H}{ii} region \object{RCW~164}.

\begin{acknowledgements}
We are grateful to J.~Casares for allowing us to mention several new parameters of LS~5039 prior to publication.
The NRAO is a facility of the National Science Foundation operated under cooperative agreement by Associated Universities, Inc. 
The European VLBI Network (http://www.evlbi.org/) is a joint facility of European, Chinese, South African, and other radio astronomy institutes funded by their national research councils.
This work made use of the Swinburne University of Technology software correlator, developed as part of the Australian Major National Research Facilities Programme and operated under licence.
This research has made use of the SIMBAD database, operated at CDS, Strasbourg, France.
This publication makes use of data products from the Two Micron All Sky Survey, which is a joint project of the University of Massachusetts and the Infrared Processing and Analysis Center/California Institute of Technology, funded by the National Aeronautics and Space Administration and the National Science Foundation.
We acknowledge support by the Spanish Ministerio de Ciencia e Innovaci\'on (MICINN) under grants AYA2010-21782-C03-01 and FPA2010-22056-C06-02.
J.M. acknowledges support by MICINN under grant BES-2008-004564.
M.R. acknowledges financial support from MICINN and European Social Funds through a \emph{Ram\'on y Cajal} fellowship.
J.M.P. acknowledges financial support from ICREA Academia.
\end{acknowledgements}

\bibliographystyle{aa}
\bibliography{art}

\begin{appendix}

\section{Pulsar gating on PSR~J1825$-$1446} \label{obs_psr_gating}

To improve the sensitivity of the detections, and thus the astrometry, we
correlated the three VLBA observations using pulsar gating
\citep{brisken02,chatterjee09}. In normal VLBI observations, with integration
times of a few seconds, one measures the period-averaged flux density of the
observed pulsar, although most of the time the pulsar is not emitting. The
instantaneous flux density during the pulsation is usually tens of times higher
than the period-averaged value, because the duty cycle (the ratio of the pulse
width to the pulse period) is usually below 10\%. To optimise radio observations
of pulsars one can take advantage of this fact by only correlating the data
during on-pulse, and disabling correlation during off-pulse. The total
correlated on-source time is considerably reduced (a few minutes for every
observed hour), which increases the final image noise (rms) by a factor
proportional to one over the square root of the duty cycle ($\Delta\tau_{\nu}$)
at a certain frequency $\nu$. However, the flux ($S_{\nu}$) increase is
approximately proportional to one over the duty cycle, yielding a final
signal-to-noise ratio ($S/N$) gain:
\begin{equation}
(S/N)_{\rm G} = \frac{S_{\nu, G}}{rms_{G}}\sim\frac{\langle S_{\nu}\rangle/\Delta\tau_{\nu}}{\langle rms\rangle/\sqrt{\Delta\tau_{\nu}}}={(S/N)}_{0}\frac{1}{\sqrt{\Delta\tau_{\nu}}},
\label{eq:period}
\end{equation}
where the subindex G stands for the gated values. This provides a theoretical
signal-to-noise ratio increase proportional to one over the square root of the
duty cycle of the pulsar. Typically, the real signal-to-noise improves by a
factor between 2 and 6 \citep{brisken02}, which can make the difference between
a non-detection and accurate astrometry. The duty cycle of PSR~J1825$-$1446 is
2.5\% (6.8~ms) at 4.8~GHz, yielding a maximum theoretical signal-to-noise gain
of 6.3. For the three VLBA correlations, updated ephemeris of the pulse time of
arrivals were obtained to configure the opening and closing time ``gates'' of
the correlator. The timing was obtained thanks to the regular pulsar monitoring
conducted in the Jodrell Bank Observatory. Two sets of data were obtained for
each observation, one from a normal correlation and one using pulsar gating.
This allowed us to compare the results obtained with and without the pulsar
gating correlation. The three VLBA observations were correlated with a gate
covering pulsar phases from 0.99056 to 0.01389, integrating 6.5~ms of each
pulsar cycle. 

The same calibration tables applied to the normal data were also applied to the
data correlated with pulsar gating. We produced images of
\object{PSR~J1825$-$1446} using the same parameters, and we compared the
astrometry obtained with and without pulsar gating. The obtained signal-to-noise
ratio and the corresponding gain factor improvement are shown in
Table~\ref{table:psr_obs}. On average, the obtained gain of 2 is relatively
small. This can be explained by an asymmetric pulse profile, or an inaccurate
determination of the pulsar gate.

\end{appendix}

\end{document}